\documentclass[a4paper,fleqn,usenatbib]{mnras}
\usepackage{newtxtext,newtxmath}
\usepackage[T1]{fontenc}

\DeclareRobustCommand{\VAN}[3]{#2}
\let\VANthebibliography\thebibliography
\def\thebibliography{\DeclareRobustCommand{\VAN}[3]{##3}\VANthebibliography}

\usepackage[dvipsnames]{xcolor}
\usepackage{ae,aecompl}
\usepackage{graphicx}	
\usepackage{amsmath}	
\usepackage{mathtools}
\usepackage{array}
\usepackage{booktabs}
\usepackage{hyperref}
\hypersetup{draft=false}

\title[DEVILS: Morphology-Density Relation]{Deep Extragalactic VIsible Legacy Survey (DEVILS):  Evolution of the Morphology-Density Relation}
\author[L. J. M. Davies]{L. J. M. Davies$^{1}$\thanks{E-mail:
 luke.j.davies@uwa.edu.au},  J. Doan$^{1}$, S. Bellstedt$^{1}$, A. S. G. Robotham$^{1}$, S. Phillipps$^{2}$, C. Wolf$^{3}$, M. Meyer$^{1}$,  \newauthor M. Siudek$^{4,5}$, S. P. Driver$^{1}$ \\
 $^{1}$ ICRAR, The University of Western Australia, 35 Stirling Highway, Crawley, WA 6009, Australia \\
 $^{2}$ Astrophysics Group, School of Physics, University of Bristol, Bristol BS8 1TL, UK \\
 $^{3}$ Research School of Astronomy and Astrophysics, Australian National University, Canberra, ACT 2611, Australia\\
 $^{4}$ Instituto de Astrof\'{\i}sica de Canarias, V\'{\i}a L\'actea, 38205 La Laguna, Tenerife, Spain \\
 $^{5}$ Instituto de Astrof\'isica de Canarias (IAC); Departamento de Astrof\'isica, Universidad de La Laguna (ULL), 38200, La Laguna, Tenerife, Spain\\
}

\date{Accepted XXX. Received YYY; in original form ZZZ}

\pubyear{2025}

\begin{document}
\label{firstpage}
\pagerange{\pageref{firstpage}--\pageref{lastpage}}
\maketitle

\begin{abstract}
Galaxies with different morphological characteristics likely have different evolutionary histories, such that understanding the mechanisms that drive morphological change can provide valuable insights into the galaxy evolution process. These mechanisms largely correlate with local environment, ultimately leading to the well-known local morphology-density relation. To explore \textit{how} the morphology-density relation is produced, we must look to earlier times, and trace the co-evolution of environment and morphology in an un-biased and self-consistent manner. Here we use new environmental metrics from the Deep Extragalactic VIsible Legacy Survey (DEVILS) to explore the spectroscopic morphology-density relation at intermediate redshift (0.3$<$\textit{z}$<$0.5) and compare directly to the Galaxy And Mass Assembly Survey (GAMA) at 0$<$\textit{z}$<$0.08. Importantly, both the galaxy morphologies and environmental metrics in DEVILS and GAMA are derived in a very similar manner, reducing any methodology biases. We see a clear evolution in morphological classes between DEVILS and GAMA, which is modulated by environment. These trends are consistent with a scenario where in all environments disk-dominated galaxies are transitioning to classical bulge+disk systems (potentially via minor mergers and/or secular evolution), and in high-density environments there is an increasing prevalence of visually-selected elliptical galaxies (potentially via major mergers and/or disk fading); with the fraction of ellipticals increasing by $\sim$0.3 in the most dense regions over the last $\sim$7\,Gyr, but remaining largely unchanged in low-density environments.                             

\end{abstract}


\section{Introduction}

Galaxies come in all sizes and shapes \citep[which we call \textit{morphology}, $e.g.$][]{Lintott08}. There are many and varied ways to define a galaxy's morphology, each with their own successes and failings \citep[$e.g.$][]{Strateva01, Kauffmann04, Conselice06, Scarlata07, Schawinski07, Driver06, Driver22}. Early studies into defining galaxy morphologies largely taxonomically-classified systems into four broad groups based on their visual appearance: i) elliptical galaxies, which have an elliptically-shaped smooth light profile, with no clearly-definable internal structures, ii) disk-dominated galaxies, which have a disk-like structure and can contain internal features such a spiral arms, central bulges, bars, etc, iii) lenticular or S0 galaxies which sit as an intermediary stage between ellipticals and disk-dominated systems, and iv) irregular systems, which have no discernible axis-symmetric structure and a clumpy light distribution \citep[these classifications have largely remained similar to the original work of][]{Hubble26}. Early works suggested that different morphological types could be indicative of different evolutionary histories, and attempted to understand the astrophysical mechanics that could lead to the distribution of morphological classes we see today \citep[$e.g.$ more recently see][]{Driver06, Bezanson09, Hopkins09, Naab09}. 

However, with the development of higher resolution telescopes, and specifically space-based telescopes such as the Hubble Space Telescope (HST), and now the James Webb Space Telescope (JWST) and Euclid Space Telescope, which avoid some of the limitations of atmospheric seeing, the morphological picture of galaxies has become more and more complicated \citep[$e.g.$ see][]{Barden05, Wolf05, Willett16, Hashemizadeh21, Company24, Lee24, Company25, Euclid25}. With increasing resolution, we observe that the morphological and structural characteristics of almost all galaxies are somewhat unique, and robustly classifying the galaxies we observe into one of the four broad camps detailed above is not as straightforward as we first thought. This is particularly true when trying to define galaxy morphology to relate to some key tenet of the galaxy's evolutionary past, leading us to subdivide the population into ever more niche morphological classes, for example separating based on diffuse (or pseudo) bulge as compared to classical bulge \citep{Graham99,Kormendy04, Gadotti09, Driver22}, bars and their strength \citep{Masters11}, clumps \citep{Elmegreen09}, asymmetry \citep{Conselice00}, structural shape parameters \citep{Scarlata07}, rings \citep{Kelvin18}, etc. All of these structures are likely, at some level, to be indicative of a galaxy's evolutionary history.  

Some subsequent studies have aimed to be less subjective, defining a galaxy's morphology using quantitatively-derived properties. These first aimed to classify a galaxy based on its surface brightness profile, largely finding that the majority of galaxies could be well-characterised by the S\'{e}rsic function \citep{Sersic63}. The dominant morphological indicator in this function is the S\'{e}rsic index (n), which defines the change in steepness of the light profile. Elliptical-like morphologies typically have a steep (n$\sim$4) profile, and disk-like morphologies show a shallower (n$\sim$2) profile. As such, one can bin galaxies into morphological classes based on fitting their light profiles to measure the S\'{e}rsic index \citep[$e.g.$][]{McIntosh05,Gadotti09, Tasca11, Kelvin12}. However, the local galaxy population shows a continuum of S\'{e}rsic indices, with only weak bi-modality between morphological types \citep[$e.g.$][]{Lange15}. Thus, this approach becomes problematic. What is more, this also assumes that all galaxies have a smooth light profile, and does not take into account any substructure within the galaxy.  

Studies with better resolution have shown that many galaxies can be better fit with two S\'{e}rsic components, one with a steep elliptical-like S\'{e}rsic index (the bulge) and one with a shallow  S\'{e}rsic index (the disk) \citep[$e.g.$][]{Allen06, Maltby12, Lange15,Casura22, Cook25}. These can be combined in varying weights to morphologically classify galaxies based on properties such as their light-weighted bulge-to-total ratio. With, to first order, pure elliptical galaxies being identified as bulge-to-total=1 systems (all bulge), pure disk galaxies as bulge-to-total=0 (no bulge), and bulge+disk systems lying in between. However,  many galaxies which require two S\'{e}rsic components, have been found to be better fit with two shallow disk-like S\'{e}rsic index profiles instead of a steep elliptical-like and disk component, leading to the structural separation of pseudo-bulges (disk-like) and classical-bulges (elliptical like) - potentially with different formation/evolutionary roots \citep{Hashemizadeh22, Cook25}. Complicating matters, the majority of these studies are performed at a single observed-frame wavelength. This is problematic as typically different structural components ($i.e$ bulges and disks) have different mass-to-light ratios, meaning that: i) by fitting in a single band we are not accurately tracing the stellar content of galaxies in a consistent manner across different components, inducing a bias \citep{Kelvin12}, and ii) if there is a redshift range across a given sample, we are not comparing similar stellar populations for different galaxies, also inducing a bias. For example, galaxies appear clumpier/more irregular when observed at bluer rest-frame wavelengths, and due to stellar age gradients, can appear larger and less concentrated. These combined mean that defining morphological classes using surface brightness light profiles is also fraught with difficulty. Recently, a number of studies have aimed to overcome this by performing spectro-structural decompositions, essentially fitting a galaxy's spectral energy distribution (SED) and structural light profile simultaneously to avoid the complications noted above \citep[$e.g$][]{Robotham22, Bellstedt24}. $i.e$ allowing a different mass-to-light ratio for each component, and fitting in the rest-frame of each galaxy. These studies are in their infancy and require a significant investment of processing times to yield results for large statistical samples.  

At higher redshifts, where structural decompositions can be problematic due to limited spatial resolution, quantitive measures of galaxy light profiles such as concentration, asymmetry and smoothness \citep[CAS parameters, $e.g$][]{Conselice03} can be used to define galaxy morphologies \cite[$e.g.$][]{Kartaltepe23}, revealing evolution in the distribution of morphological types in the very early Universe. These types of approaches can also be extended to modern machine learning techniques to automatically classify galaxies in large area surveys \citep[$e.g.$][]{Cavanagh21, Vavilova21, Tarsitano22}. While these studies allow efficient processing of huge galaxy samples, care must be taken as they still rely on robust training samples and significant post-processing data validation.            

Finally, a potentially more-robust approach when aiming to link a galaxy's structure to its evolutionary history, is to use the dynamics of galaxy components instead of their visual appearance. Processes such as mergers, tidal interactions, harassment, etc will likely leave stronger imprints on a galaxy's stellar and gas dynamics, than on its visible light profile. This approach has been used to great effect in large surveys using integral field spectroscopy (IFS) to quantify galaxy morphologies using their dynamics and identify the distribution of different morphological types \citep[$e.g.$][]{Cortese16, Wang20, Watson22, Vaughan24}. However, these observations are costly to obtain as high signal-to-noise is required to robustly map galaxy dynamics, and are usually only available at ground-based facilities with somewhat poor spatial resolution. This can limit such studies to just massive galaxies in the very local universe and/or small sub-samples of the galaxy population. Interestingly, these results have revealed a potential tension between visual morphological classifications and those defined based on galaxy dynamics - with common visually-identified morphological types showing a broad range of dynamics, suggesting diverse evolutionary histories. This is especially true for high stellar mass elliptical galaxies, where only $\sim50\%$ of early-type galaxies are found to show little or no dynamically rotating component \cite[$e.g.$ as would be expected from a \textit{true} elliptical-like system,][]{Guo20}. Potentially the remaining systems were originally two-component, bulge+disk systems, whose disks have faded and are now visually identified as elliptical, but whose dynamics retain a rotational component.   These results are also consistent with more recent studies exploring the dynamically-defined morphological classifications \citep[$e.g.$][]{FraserMcKelvie22, Rigamonti24} - indicating that care must be taken when linking a galaxy's visual morphological class to potential evolutionary histories. Exploring dynamically-defined morphological classifications in comparison to visual classifications for large samples of galaxies outside of the very local Universe, is currently not possible \citep[however, see recent attempts by][]{Gillman20, Amvrosiadis25, Espejo25}.                       

\vspace{2mm}

As one can see, the morphological classification of galaxies, and inferences therein about galaxy evolution is a complicated and messy field. Despite all of these problems, careful and measured studies into the morphological classification of well-defined samples of galaxies have led to invaluable insights into the galaxy evolution process. We have now robustly mapped out the morphological distribution of galaxies in terms of stellar mass in the local Universe, through large surveys such as the Sloan Digital Sky Survey,  SDSS, \citep{Weigel16} and Galaxy And Mass Assembly Survey, GAMA, \citep{Moffett16, Driver22}, and have also begun to robustly probe the global evolution of morphological types \citep{Vulcani11, Hashemizadeh21, Kolesnikov25}. These observations, combined with theoretical models for galaxy evolution, suggest that compact bulge-like structures may form in the early Universe \citep{Chen22, Benton24, Bodansky25},  which then grow disks \citep[likely through gas accretion, $e.g.$][]{Ho19} and larger bulges \citep[potentially growing larger through minor mergers, $e.g.$][]{Zavala12}, these systems then likely combine in major mergers, ultimately forming lenticular and then elliptical galaxies \citep{GonzalezGarcia05} as the end point of the structural evolution process  \citep[$e.g.$ see][]{Driver06}. This can also be seen via complementary approaches exploring the star-formation histories of galaxy components in the local Universe, finding that compact bulges form very early, with the other components accumulating over universal history \citep[$e.g$][]{Bellstedt24}. However, the evolutionary mechanisms which lead to the morphological distribution of galaxies we see today is far from clear, and there is currently much debate as to both the primary mechanism by which this occurs and the frequency/importance of different evolutionary processes.   

However, as we now have some idea of the global morphological distribution of galaxies and its evolution, the question we must now ask is why? What astrophysical processes drive these morphological changes and to what degree do they affect different subsamples of the galaxy populations? The evolutionary models suggested above are largely based on theoretical interpretations of limited data. However, can we observationally identify the processes which lead to morphological change? The most-notable early observation in this area, was the identification that there is a strong correlation between a galaxy's local environment and its morphological type \citep[the morphology-density relation, $e.g.$][]{Hubble31,Oemler74, Davis76, Dressler80, Postman84}. These works showed that elliptical galaxies are more prevalent in high-density environments, while the converse is true for disk-dominated galaxies, and suggested that there was some aspect of the over-dense environment that facilitates the transition of galaxies from disk-dominated to lenticular and elliptical morphologies, over and above the transformations that were occurring in low-density environments. $i.e.$ that over-dense environments were accelerating morphological evolution. Galaxy interactions and major mergers were the proposed mechanism for these transitions, and can today be witnessed occurring in over-dense environments with great frequency \citep[particularly in group-scale environments, where galaxy volume densities are high, but velocity dispersions are relatively low, $e.g.$see][]{Pearson24}. However, this picture is not so clear cut. Other studies have suggested that, particularly for ellipticals, the morphology density relation is only a consequence of more massive galaxies existing in more over-dense environments, and that elliptical morphology is strongly correlated with stellar mass \citep[][]{Bamford09, Alpaslan15,vanderWel10, Pfeffer23}. They argue that, yes, morphological transitions from disk-dominated to lenticular galaxies do occur in high-density environments - partially leading to the morphology-density relation - but that for ellipticals there is no accelerated morphological transition in cluster environments.   

Today, the local morphology-density relation has been mapped out in ever increasing detail, robustly defining the \textit{correlation} between a galaxy's visual appearance and local environment \citep[$e.g.$][]{Houghton15}, and kinematic morphology and environment \cite[$e.g.$][]{Cappellari11, Fogarty14,Greene17}. These trends have also been observed in modern hydrodynamic simulations, suggesting they are a ubiquitous feature of the galaxy evolutions process \citep[$e.g.$][]{Pfeffer23}. However, while we know the exquisite details of these \textit{correlations} in the local Universe, their \textit{causation} at earlier times is tricker to directly observe. Many studies have aimed to map the evolution of the morphology-density relation over cosmological timescales; to witness the build-up of the $z$$\sim$0 morphology-density relation in action \citep[$e.g.$][]{Dressler97, Smith05, Postman05, vanDerWel07, Tasca09, Siudek22, Cleland25}. These studies have resulted in varying degrees of success, but overall find that we can see evidence of an increasing elliptical+lenticular fraction in the most over-dense environments - likely observing the build-up of the correlation between ellipticals and high-density environments. However, as noted above, robustly and consistently measuring galaxy morphology over a broad redshift range is fraught with difficulty, potentially leading to significant biases in these trends. What's more, even defining the local galaxy density can be problematic and vary significantly as a function of a particular study. This is largely due to the fact that outside of the local Universe there has been a paucity of high-completeness spectroscopic samples with which to define galaxy environments. This leads studies to rely on, sometimes low-precision,  photometric redshifts to define the galaxy density. Given the precision of photometric redshifts, chance alignments of galaxies (or the converse) can lead to biases in measurements of the local galaxy density, which can also vary as the function of the specific redshifts probed.  To further complicate the matter, trying to compare the morphological evolution of galaxies across multiple epochs, and linking this with the co-evolution of environment strongly relies on consistency in the methodology employed for both morphological classifications and environmental definitions across the full redshift range. Thus, comparing results from different teams using different data and methodologies is extremely problematic. 

Here we aim to overcome some of these issues by using a combination of the Deep Extragalactic VIsible Legacy Survey \citep{Davies18, Davies21} and Galaxy And Mass Assembly Survey \citep[GAMA][]{Driver11, Driver16, Liske15, Baldry18}. These surveys both provide deep, high spectroscopic completeness samples with which to study galaxy and environmental evolution. GAMA provides a large area sample in the local Universe ($z$$<$0.2), while DEVILS provides a comparable, small area, sample extending out to $z$$<$1. Importantly, DEVILS was designed to match GAMA in a number of respects, but to extend out to the more distant Universe - allowing us to define comparable samples of galaxies across a broad evolutionary range.  Within these samples we now also have galaxy properties (stellar masses, morphologies, etc) and environmental diagnostics (nearest-neighbour density, etc) derived by the same teams in an identical manner using the same codes and techniques. This allows us to probe galaxy evolution process in both surveys simultaneously, while minimising methodology biases and inconsistencies. In this work we will use the DEVILS and GAMA samples in combination to explore the evolution of the morphology-density relation over the last $\sim$5\,Gyr ($i.e.$ $z$$<$0.5) using a stellar mass limited and fully spectroscopic sample.  Throughout this paper we use a Planck cosmology \citep{Planck20} with {H}$_{0}$\,=\,68.4\,kms$^{-1}$\,Mpc$^{-1}$, $\Omega_{\Lambda}$\,=\,0.699 and $\Omega_{M}$\,=\,0.301.

\section{Data and Sample Selection}

In this work we use samples derived from both the Deep Extragalactic VIsible Legacy Survey (DEVILS) at intermediate redshift and the Galaxy And Mass Assembly Survey (GAMA) at low redshift. While the data products used in this work are described extensively elsewhere, we briefly summarise the key measurements here.

\subsection{Note on Morphological Classifications}

First, as noted above, the morphological classification of galaxies is problematic, especially when aiming to link galaxy morphologies to different evolutionary histories. This is particularly true when trying to map the time evolution of morphological classes when different observations/techniques are employed at different epochs. In this work we use the arguably most straightforward, but potentially most subjective, morphological classification from human visual inspection of deep imaging surveys. While these are potentially biased, and may mis-classify some galaxies when aiming to identify common galaxy components (as described above, $i.e.$ the difference between visually-selected and dynamically-selected morphologies), this does allow us to employ very similar classification techniques over a broad range of look back times, where many other techniques are limited to a single epoch. We also aim to limit any bias in our visual morphological classifications by having multiple classifiers and a multi-stage classification process, and use very similar (rest-frame) data to perform these classification across all epochs probed. We would therefore argue that, yes, visual morphological classifications can be biased, but that the biases induced by this methodology are largely the same across all of the epoch probed in this work - allowing us to explore the time evolution of morphological classes. These classifications are described in more detail below.                

\subsection{Data}

\subsubsection{The Deep Extragalactic VIsible Legacy Survey}

DEVILS is a spectroscopic survey undertaken at the Anglo-Australian Telescope (AAT), which aimed to build a high completeness ($>$85\%) sample of $\sim$50,000 galaxies to Y$<$21\,mag in three well-studied deep extragalactic fields: D10 (COSMOS), D02 (ECDFS) and D03 (XMM-LSS). The survey aims to provide the first high completeness sample at $0.3<z<1.0$, allowing for the robust parametrisation of group and pair environments in the distant Universe.  DEVILS also serves as a precursor to the Wide Area VISTA Extragalactic Survey \citep[WAVES,][]{Driver19} deep program, which will cover similar galaxy populations, and use similar methodologies to derive galaxy and environmental properties, but will cover $\sim$15 times the volume.  The science goals of both DEVILS and WAVES deep are varied, from the environmental impact on galaxy evolution at intermediate redshift, to the evolution of the halo mass function over the last $\sim$7\,billion years. For full details of the DEVILS survey science goals, survey design, target selection, photometry and spectroscopic observations see \cite{Davies18} and \cite{Davies21}. Full details of the DEVILS spectroscopic sample will be presented in Davies et al (in prep).
   
The DEVILS regions were chosen to cover areas with extensive existing and ongoing imaging to facilitate a broad range of science. In this work we only use the DEVILS D10 field which covers the Cosmic Evolution Survey region \citep[COSMOS,][]{Scoville07}, extending over 1.5deg$^{2}$ of the UltraVISTA \citep{McCracken12} field and centred at R.A.=150.04, Dec=2.22. This field is prioritised for early science as it is the most spectroscopically complete, has the most extensive multi-wavelength coverage of the DEVILS fields,  has already been processed to derive robust galaxy properties through spectral energy distribution (SED) fitting, and has had robust environmental metrics defined (see below).

\begin{figure}
\begin{center}
\includegraphics[scale=0.72]{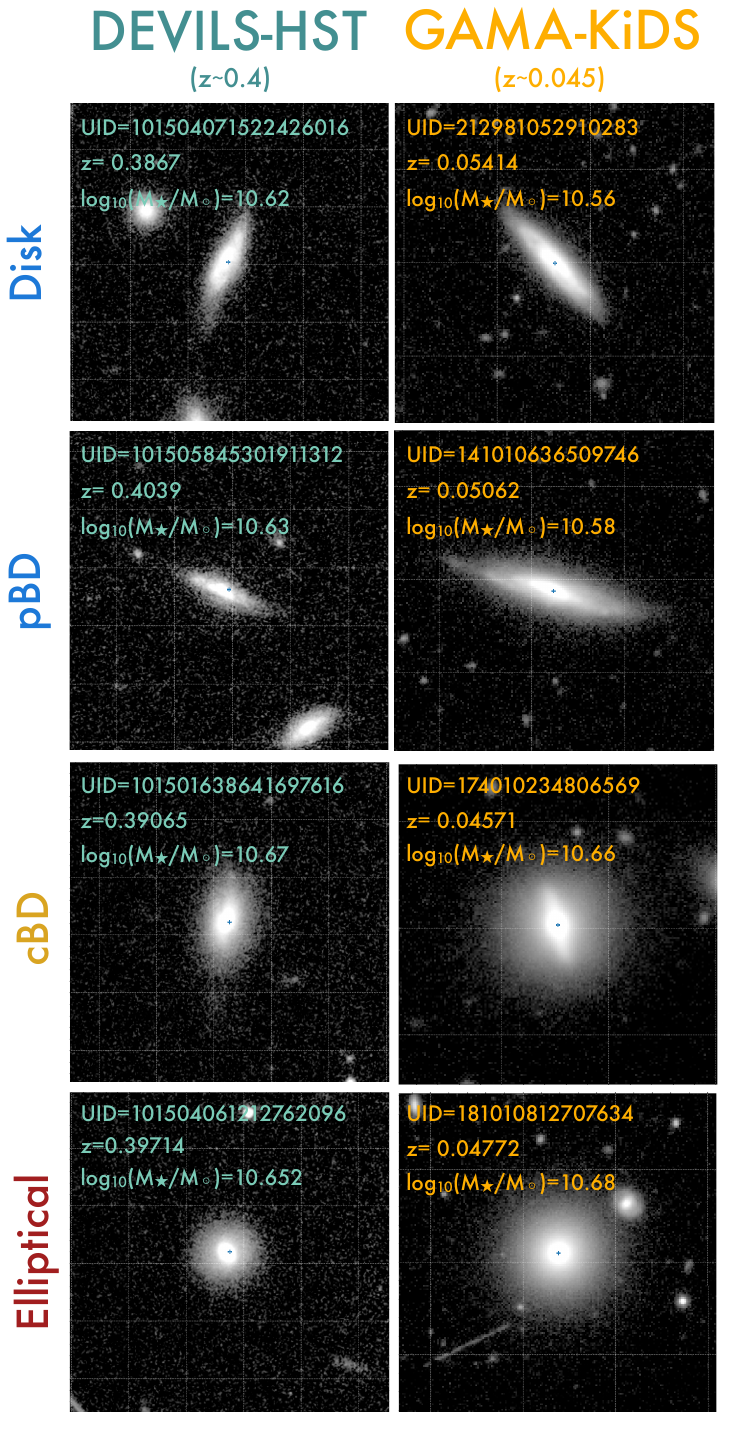}

\caption{Example of images used in the visual morphological classification of galaxies in DEVILS (left column) and GAMA (right column) . Here we select galaxies at log$_{10}$(M$_{\star}$/M$_{\odot}$)$\sim$10.6 and at $z$$\sim$0.4 for DEVILS and $z$$\sim$0.045 for GAMA, and then randomly select a galaxy in each morphological type (rows). DEVILS shows HST-ACS F814W images covering a 5\,arcsec radius cutout ($\sim$27\,kpc at $z$$\sim$0.4), while GAMA shows KiDS-r images covering a 30\,arcsec radius cutout ($\sim$27\,kpc at $z$$\sim$0.045). The blue crosses show the DEVILS/GAMA target sources positions. For DEVILS these are defined from lower-resolution VISTA imaging (and are slight offset from the HST centres).}
\label{fig:images}
\end{center}
\end{figure}

Within this work, we first use stellar masses from the SED fitting described in \cite{Thorne21} and \cite{Thorne22}. \cite{Thorne21} use the \textsc{ProSpect} \citep{Robotham20} SED fitting code to estimate galaxy properties such as stellar mass, SFR and star-formation history (SFH). In \cite{Thorne22}, this process is updated to include an Active Galactic Nuclei (AGN) model, which allows for the identification of sources hosting bright AGN and improvements to the other derived properties for AGN host galaxies. The \textsc{ProSpect} analyses of \cite{Thorne22} uses a parametric skewed-log-Normal truncated SFH for all galaxies, which assumes that all SFHs are smoothly evolving with lookback time. The use of these SFHs and a detailed description of their merits/failings are discussed extensively in \cite{Robotham20, Bellstedt20a, Thorne21, Thorne22, Bravo22, Bravo23, Lagos24,Davies25a,Davies25b} and others. As such, we do not discuss it here. In this work we use the best-fit stellar mass taken from the \textsc{ProSpect} fits which contain a possible AGN component (which likely have better-defined properties).

For visual morphologies we use the classifications outlined in \cite{Hashemizadeh21} who use HST-ACS imaging from the COSMOS survey \citep{Scoville07} to visually classify $\sim$36,000 galaxies in the D10 region. However, we briefly note that galaxies in the morphological sample are selected to cover all log$_{10}$(M$_{\star}$/M$_{\odot}$)$>$9.5 and $z$$<$1.0 systems in the overlap region between COSMOS and D10. Outside of this range of stellar mass even HST resolution imaging does not have adequate spatial resolution to visually constrain galaxy morphology, and above this redshift the traditional picture of galaxy morphological classes begins to break down as more and more galaxies become clumpy and asymmetric (both caused by true structural evolution and observational effects from observations probing bluer rest-frame emission). Galaxies within the sample limits are then classified as either Elliptical, pure disk, disk + diffuse/pseudo bulge (hereafter pDB), disk + classical bulge (hereafter, cBD),  compact (too visually small to classify) and hard (asymmetric, merging, clumpy, extremely compact, and low-S/N systems) based on their F814W HST imaging. Galaxies were independently assessed by 5 classifiers, and objects with three or more votes in one category were adopted. Where classifiers disagreed, classes were debated until a consensus was obtained. Details of potential biases and systematics in this classification are described in \cite{Hashemizadeh21}. In this work we use sources identified as elliptical, pure disk, pDB and cBD, ignoring all other classes. Subsequent work using this sample has suggested commonalities between the characteristics of the disk and pBD system, potentially suggesting that pBD galaxies are secularly-evolved later stage of pure disk systems, where the diffuse/pseudo bulge is formed through gas accretion and/or disk instabilities \cite[see][]{Devergne20, Hashemizadeh22, Cook25}. As such, in our subsequent analysis we combine disk and pBD systems into a single `disk + pBD' class. This essentially contains all disk-dominated systems.

\begin{figure*}
\begin{center}
\includegraphics[scale=0.7]{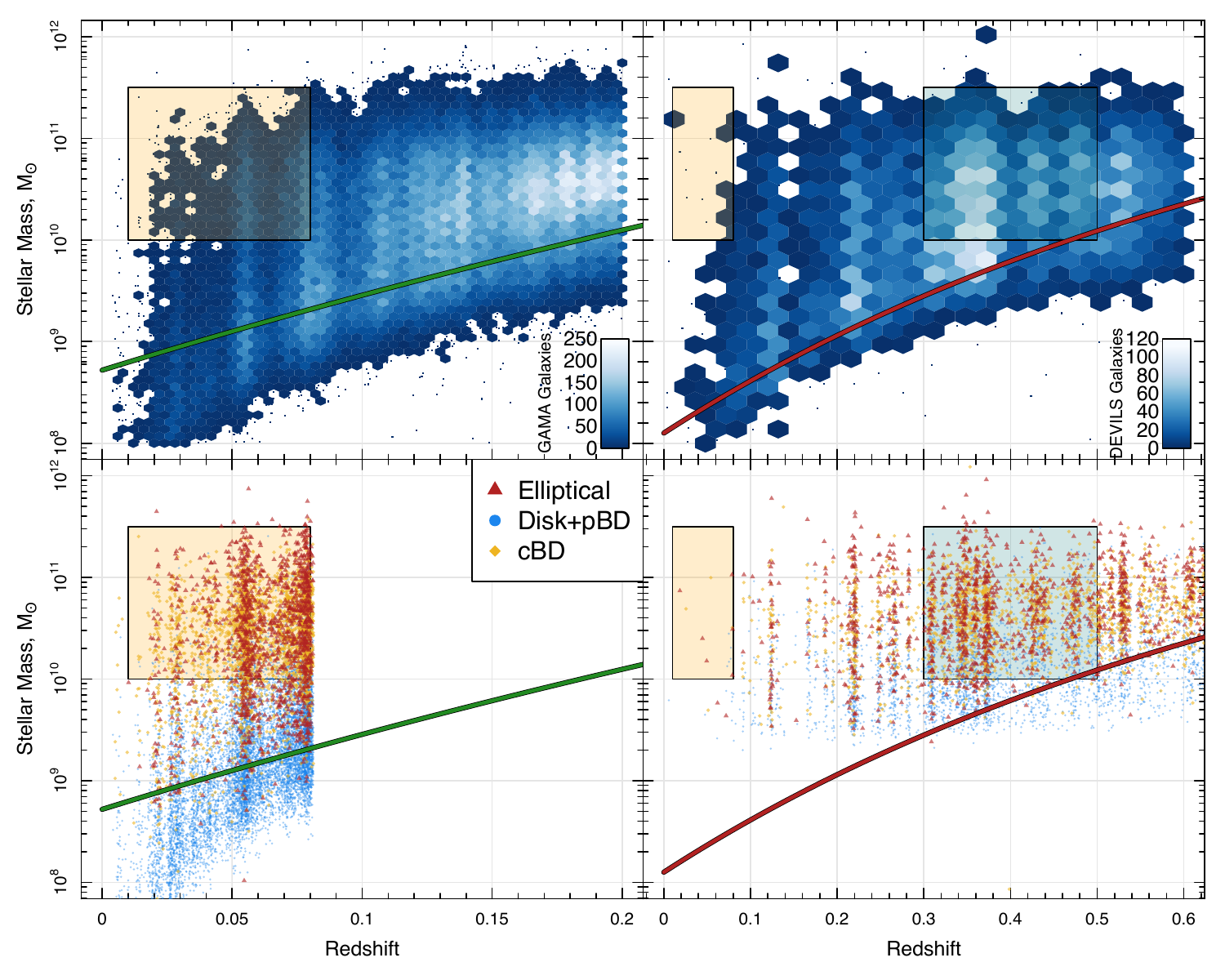}

\caption{The redshift-mass distribution of all galaxies (top row) in GAMA (left column) and DEVILS (right column) with spectroscopically-measured redshifts. Green and red lines show the stellar completeness limits for GAMA and DEVILS respectively. The bottom row shows the same, but with point type/colour showing available visual morphological classifications. Importantly, all properties (stellar mass, redshift, morphology) are derived using the same method in both GAMA and DEVILS - allowing a minimally-biased comparison across epochs.  Stellar mass and redshift ranges used in this work are shown in coloured polygons. DEVILS at 0.3$<$$z$$<$0.5 provides a similar sample in terms of stellar mass to GAMA at $z$$<$0.08, allowing a direct comparison of the time evolution of the morphology-density relation since $z$$\sim$0.5. }
\label{fig:sample}
\end{center}
\end{figure*}

\subsubsection{The Galaxy And Mass Assembly Survey}

The GAMA survey fourth data release (GAMA II) covers 250\,deg$^{2}$ to a main survey limit of $r_{\mathrm{AB}}<19.8$\,mag in three equatorial (G09, G12 and G15) and two southern (G02 and G23 - survey limit of $i_{\mathrm{AB}}$$<$19.2\,mag in G23) regions. The spectroscopic survey was undertaken using the AAOmega fibre-fed spectrograph \citep[][]{Saunders04,Sharp06} in conjunction with the Two-degree Field \citep[2dF,][]{Lewis02} positioner on the Anglo-Australian Telescope, and obtained redshifts for $\sim$240,000 targets covering 0$<$$z$$\lesssim$0.5 with a median redshift of $z$$\sim$0.2, and highly uniform spatial completeness \citep[see][for a summary of GAMA observations]{Baldry10,Robotham10,Driver11}. Full details of the GAMA survey can be found in \citet{Driver11, Driver16, Driver22},  \citet{Liske15} and \citet{Baldry18}. In this work we first limit our sample to galaxies that have a confirmed, local-flow-corrected redshift at 0.01$<$$z$$<$0.2.

For GAMA stellar masses, we also use the outputs of the \textsc{ProSpect} SED fitting process undertaken in an identical manner to DEVILS. This process is initially outlined in \cite{Bellstedt20b} but was subsequently expanded to include an AGN model, with the same SFH choices and methodology choices outline in \cite{Thorne22}. The GAMA  \textsc{ProSpect} fitting is undertaken using photometry covering similar rest-frame emission wavelengths as DEVILS, with photometric data outlined in \cite{Bellstedt20a, Bellstedt20b}. From this work we use best-fit stellar mass in an identical manner to the DEVILS sample described above. As such, this likely removes any significant biases induced by choice of methodology in the derived galaxy properties. The only small caveat to this is the fact that GAMA galaxies exist at a different median redshift to DEVILS (and as such have slightly different rest-frame coverage) and have different depth input photometry (and as such may have larger errors in the fitting). However, these will likely result in very minor differences in the efficacy of the SED fitting.  

For visual morphologies we use the classifications outlined in \cite{Driver22}, who use the ESO VST Kilo-degree Survey (KiDS) Data Release 4 \citep{Kuijken19} $grZ$ 3-colour imaging to visually classify 15,330, $r$$<$19.65 and $z$$<$0.08 GAMA galaxies. This redshift range is selected as the limit to which KiDS resolution imaging can be used to estimate galaxy morphologies. Importantly, visual morphological classifications are performed in a very similar manner to \cite{Hashemizadeh21} using similar classes and are undertaken by the same classifiers. The only exception is that the \cite{Driver22} work also defines a `Hard elliptical (HE)' class, but following \cite{Driver22}, in our subsequent analysis we combine these with the elliptical class. In addition, as above we combine pure disk and disk + diffuse bulge systems into a single `disk + pBD' class. For further details see \cite{Driver22}.  

\subsection{GAMA and DEVILS Visual Morphologies}
\label{sec:compVis}

We first note here that the physical spatial resolution of the KiDS imaging over the GAMA morphological samples, is comparable to the physical spatial resolution of the HST imaging in DEVILS, allowing for direct comparison of the morphological classes. For example, the 0.7\,arcsec pixel scale of KiDS equates to $\sim$0.6 physical kpc at $z$=0.045, while the 0.05\,arcsec pixel scale of HST-ACS COMSOS imaging equates to $\sim$0.3 physical kpc at $z$=0.4. To highlight this, Figure \ref{fig:images} shows a comparison of DEVILS HST-F814W imaging for $\sim$M$^{*}$, $i.e.$ log$_{10}$(M$_{\star}$/M$_{\odot}$)$\sim$10.6,  galaxies at $z\sim0.4$ and GAMA KiDS-r imaging (similar rest-frame wavelengths) for $\sim$M$^{*}$ galaxies at $z\sim0.045$ - displayed over similar physical scales. Here we show a random example of the four different visual morphological classes used in this work. While the physical spatial pixel scale of HST-ACS is $\sim\times2$\,higher for the DEVILS sample than in GAMA-KiDS, comparable visual morphological features can be observed for both samples. 

\textcolor{black}{One other potential source of bias in comparing morphological classifications across the DEVILS-HST and GAMA-KiDS imaging, is that fact that these imaging data sets have different surface brightness limits. For example, the GAMA KiDS r-band imaging reaches a 1$\sigma$ surface brightness limit of 26.25\,mag\,arcsec$^{-2}$ \citep{Bellstedt20a} while the DEVILS-HST F814W imaging only reaches a 1$\sigma$ surface brightness limit of 25.5\,mag\,arcsec$^{-2}$. This is compounded by cosmological dimming where surface brightnesses decline as $(1+z)^{-4}$.  As such, fainter galactic features may be more easily identified in GAMA-KiDS, potentially leading to the mis-classification of sources. However, in this paper we ultimately only use sources with 10$<$log$_{10}$(M$_{\star}$/M$_{\odot}$)$<$11.5 (see following section), which restricts the DEVILS-HST sample to galaxies which are $>$2magnitudes brighter than the HST extended source detection limit \cite[see][]{Scoville07} for all sources used in this paper (and $>4$ magnitudes for the highly-robust spectroscopic sample defined in the following section). As such, we are not using visual classifications close to the limiting depth of the data. In addition, these surface brightness effects likely only impact the faint outskirts of these systems, which do not dominate the morphological classifications (especially for sources far above the detection limit). }

\textcolor{black}{However, to test this further we utilise the publicly-available JWST COSMOS-Web imaging \citep{Casey23} which covers a sub-region of the COSMOS HST field. The COSMOS-Web imaging extends to fainter magnitudes than the HST imaging and therefore allows us to asses the impact of surface brightness limits on our morphological classifications. Here we use the F115W filter, as it is closest in wavelength to the HST F814W band (however, note that this may introduce some biases when comparing classifications based off different rest-frame wavelengths).  While we do not reproduce all visual morphological classifications for DEVILS sources using COSMOS-Web (which is beyond the scope of this work), we take a random sample of 500 galaxies (spanning the redshifts used in this work), and make similar cutouts to our HST sample used for the visual classifications in \cite{Hashemizadeh21}. We then blindly repeat our visual morphological classifications using the same classification classes as \cite{Hashemizadeh21}. We find that only a few percent of sources change visual classification when using the JWST data. These are almost all changes between the pBD and cBD classes, which are arguably the most subjective of visual classification categories. When compared side-by side, we see very little morphological differences between the HST and JWST imaging. Examples of this comparison a shown in Appendix \ref{app:JWST}.  As such, we argue that the differences in surface brightness limits between GAMA-KiDS and DEVILS-HST do not significantly impact our morphological classifications. }  In combination these caveats highlight that any potential biases in morphological classification between the two samples, will likely be minimal.       

\vspace{2mm}

Before proceeding, it is also worth highlighting how these visual morphological classifications map onto the more traditional classifications of, $e.g$, \cite{Hubble26}. While for single-component  pure-disk and elliptical systems these retain the similar classifications to the traditional Hubble tuning fork, for two-component systems our classification differs. Traditionally, these would have been split into all bulge+disk systems and lenticular, or S0, galaxies. However, here we opt to split 2-component systems into disk + diffuse/pseudo bulge and disk + compact/classical bulge based on their visual characteristics. This is motivated by the fact that there is growing observational evidence that bulges can be split into two evolutionary pathways \citep{Bellstedt24},  loosely divided based on their diffuse or compact morphology \cite[$e.g$][]{Cook25}. To first order diffuse/pseudo bulges potentially grow via gas accretion/disk instabilities and retain some disk-like characteristics ($i.e.$ have an $n$=2 light profile and are rotation supported), while compact/classical bulges potentially form in the early Universe through mergers and then subsequently grow a disk, and/or form later through numerous minor mergers, and have spheroidal-like characteristics ($i.e.$ have an $n$=4 light profile and are dispersion supported). As we ultimately wish to understand the evolutionary mechanisms which lead to morphological change, we aim to sub-divide galaxies based on structures with common evolutionary histories. We note that in this classification scheme, S0/lenticular galaxies typically fall in the disk + compact/classical bulge class, but also highlight that they are likely the most subjective visually-classified morphological type.     

\begin{figure*}
\begin{center}
\includegraphics[scale=0.5]{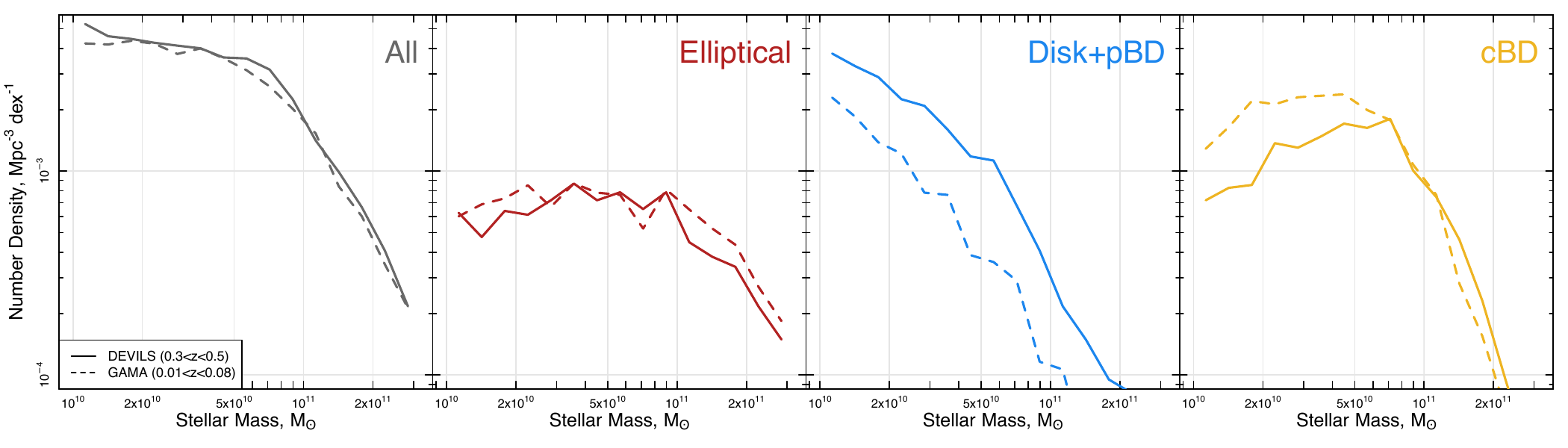}
\includegraphics[scale=0.5]{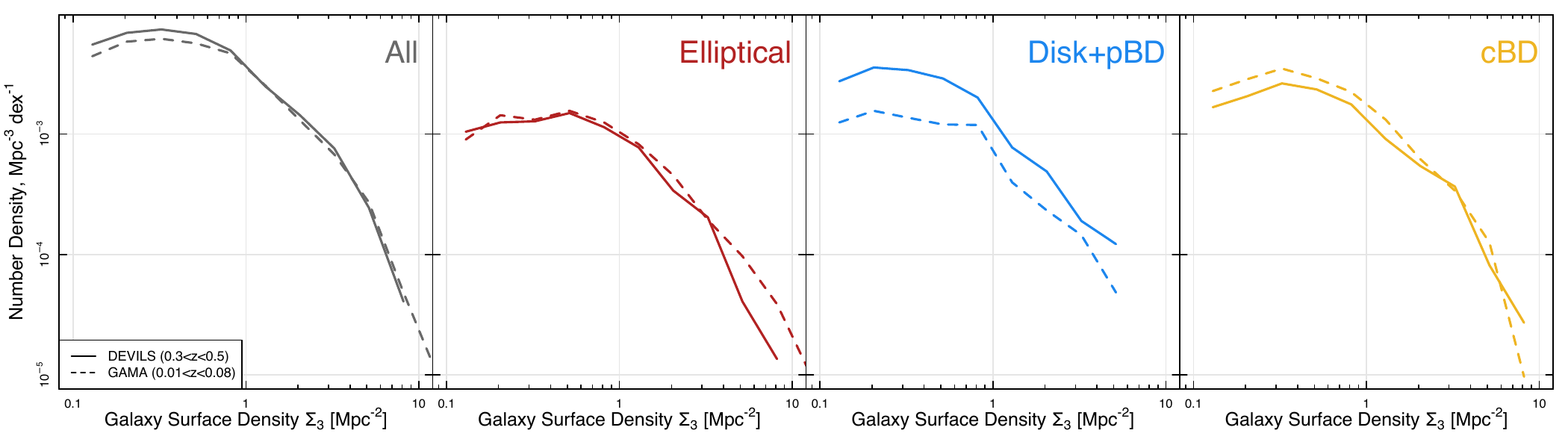}

\caption{Top row: The number density of galaxies as a function of stellar mass (the stellar mass function), split by morphological type for the samples used in this work. Solid lines show DEVILS at $0.3<z<0.5$ and dashed lines show GAMA at $z<0.08$. Bottom row:  The number density of galaxies as a function of 3rd nearest neighbour density ($\Sigma_{3}$).    }
\label{fig:hists}
\end{center}
\end{figure*}

\subsection{Sample selection}
\label{sec:select}

Using the DEVILS and GAMA samples defined above, we now wish to select subsamples in both stellar mass-redshift and morphology-redshift, which can be directly compared across multiple epochs. Figure \ref{fig:sample} displays the full sample of GAMA (left column) and DEVILS (right column) spectroscopically confirmed galaxies used in this analysis as a function of redshift. The top row shows the full stellar mass distribution, while the bottom row shows the sample, but coloured by visual morphological classes.   

First, to define a stellar mass complete sample, we use a similar approach outlined in previous works for GAMA \citep[$e.g.$][]{Davies16a} and DEVILS \citep[$e.g.$][]{Thorne21, Fuentealba25, Davies25c}. In bins of lookback time, we determine the distribution rest-frame $g-i$ colour of galaxies and find the stellar mass limit which encompasses $90\%$ of the colour distribution at a given lookback time. We then linearly fit the stellar mass limit vs look back time relation to obtain a smoothly evolving stellar mass limit with redshift. This is undertaken on GAMA and DEVILS separately and is displayed in Figure \ref{fig:sample} as the green and red lines respectively. Above these lines our samples are largely complete in stellar mass to both red and blue galaxies. We note that these completeness limits are consistent with previous GAMA and DEVILS papers. In the bottom row of Figure \ref{fig:sample}, we then show the same but only displaying points with visual morphological classifications in GAMA (left) and DEVILS (right), with different points/colours for each class.  Using the stellar mass completeness limits, and distribution of morphological classes, we then define a spectroscopic-redshift, stellar mass and morphologically complete sample of 10$<$log$_{10}$(M$_{\star}$/M$_{\odot}$)$<$11.5 galaxies for GAMA at $z$$<$0.08 and DEVILS at 0.3$<$$z$$<$0.5. These are shown as the orange and green rectangles in Figure \ref{fig:sample} respectively. This also results in survey volumes, which are roughly comparable at each epoch (1033050\, comoving Mpc$^{3}$ in GAMA and 736664\,comoving Mpc$^{3}$ in DEVILS). Note that in Section \ref{sec:highz} we extend this analysis to higher redshifts using a combination of spectroscopic and photometric redshifts. We opt to separate out these analyses, as the latter is potentially less robust and subject to additional biases.       

The aim of this work, in terms of improvement over existing studies, is to provide a highly-robust measurement of the evolution of the morphology-density relation using well-defined and minimally-biased samples across GAMA and DEVILS. By selecting the volumes, which are complete in stellar mass-redshift and morphology-redshift, by using similar visual morphological classification and (as described in the following section) identical definitions of local galaxy density using only spectroscopic redshifts, we aim to produce this minimally-biased sample spanning the last $\sim$5\,Gyrs of Universal history. 

To further justify that this is the case, the top panel of Figure \ref{fig:hists} then shows the resultant stellar mass volume number density of morphological types for both DEVILS (solid lines) and GAMA (dashed lines). First, and encouragingly, we find that the \textit{shape} of the stellar mass distributions for different morphological types are very similar across DEVILS and GAMA. This might not be too surprising given that the morphological classifications were completed by the same teams of classifiers. However, it also indicates that there are likely no strong systematic biases in our visual classifications, even though they use different input imaging and span different epochs. We note here, that this is not meant to be a detailed analysis of the morphologically-defined stellar mass function, which is described elsewhere for both GAMA \citep{Driver22} and DEVILS \citep{Hashemizadeh21}, but simply a direct comparison of the samples used in this work. However, here we do also start to see evidence of morphological evolution between the DEVILS and GAMA samples. While the overall stellar mass function (left-most panel) evolves little over this epoch \citep[consistent with][]{Hashemizadeh21}, the number density of disk \& pseudo-bulge+disk galaxies decreases with time, and the number density of classical-bulge+disk galaxies increase with time. This is also consistent with the findings \citep{Hashemizadeh21} and will be discussed later in this work.

  \begin{figure}
\begin{center}
\includegraphics[scale=0.55]{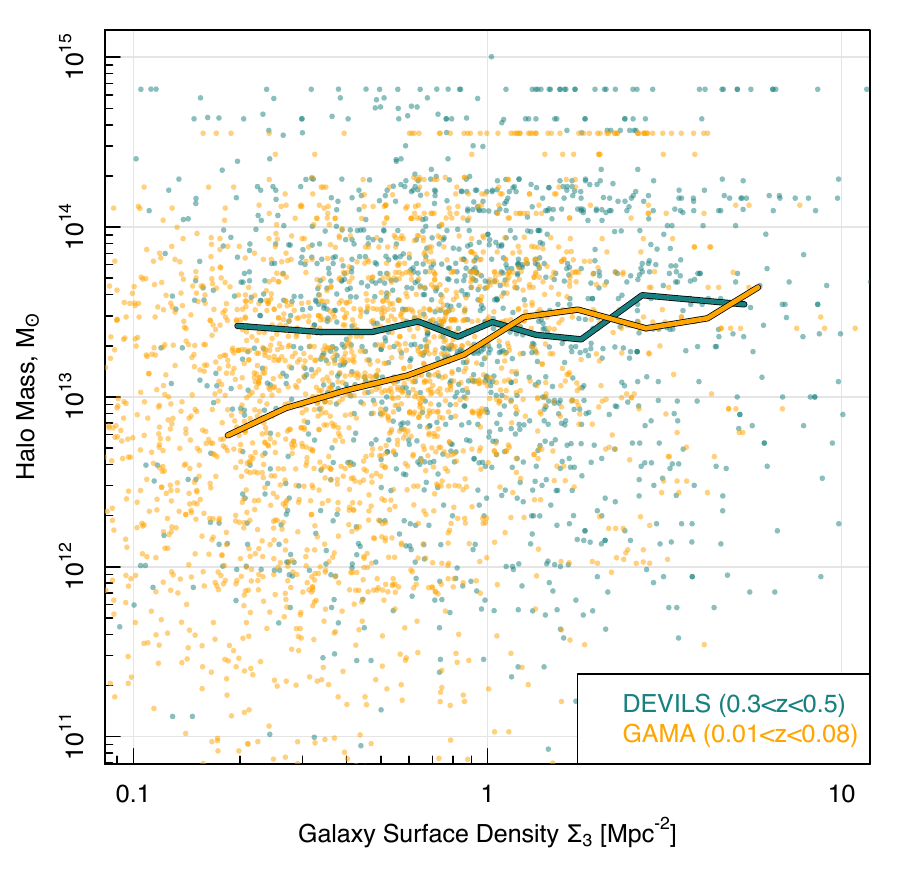}

\caption{Comparison of local galaxy density, $\Sigma_{3}$, and host dark matter halo mass for the subsample for galaxies in GAMA and DEVILS with halo mass measurements. The coloured lines display the running median of  log$_{10}$(M$_{\mathrm{halo}}$/M$_{\odot}$) }
\label{fig:halo}
\end{center}
\end{figure}

  \begin{figure*}
\begin{center}
\includegraphics[scale=0.47]{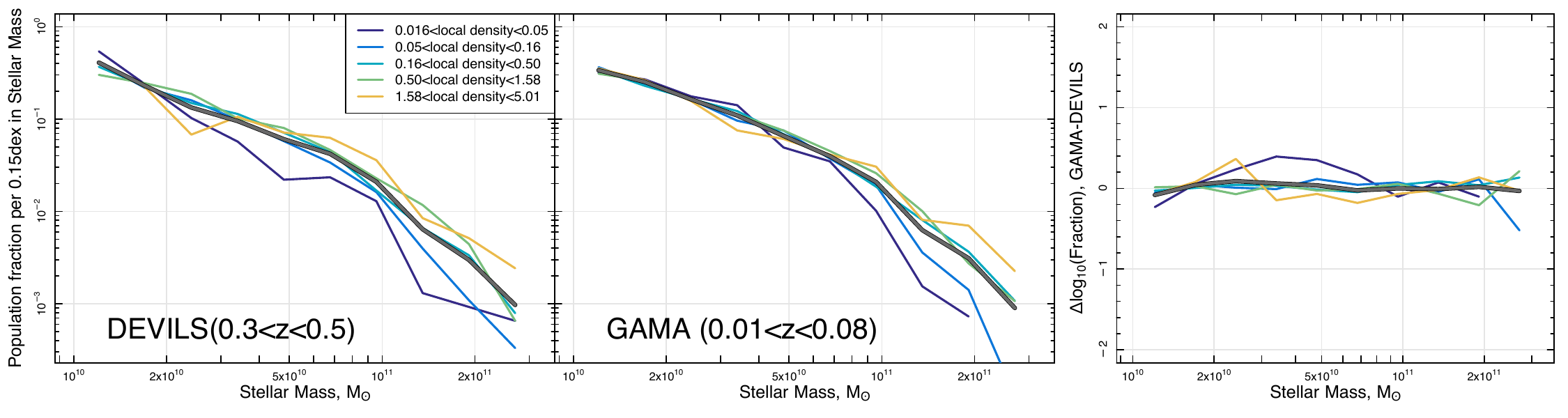}

\caption{The stellar mass population fraction histogram for different local density ranges ($\Sigma_{3}$) for DEVILS (left) and GAMA (middle). The median line is shown in grey.  The right panel shows the difference between GAMA and DEVILS at each local density range, as a function of stellar mass. This suggests there is little change in the stellar mass distribution as a function of environment between GAMA and DEVILS.     }
\label{fig:MassFuncSig}
\end{center}
\end{figure*}

\begin{figure}
\begin{center}
\includegraphics[scale=0.55]{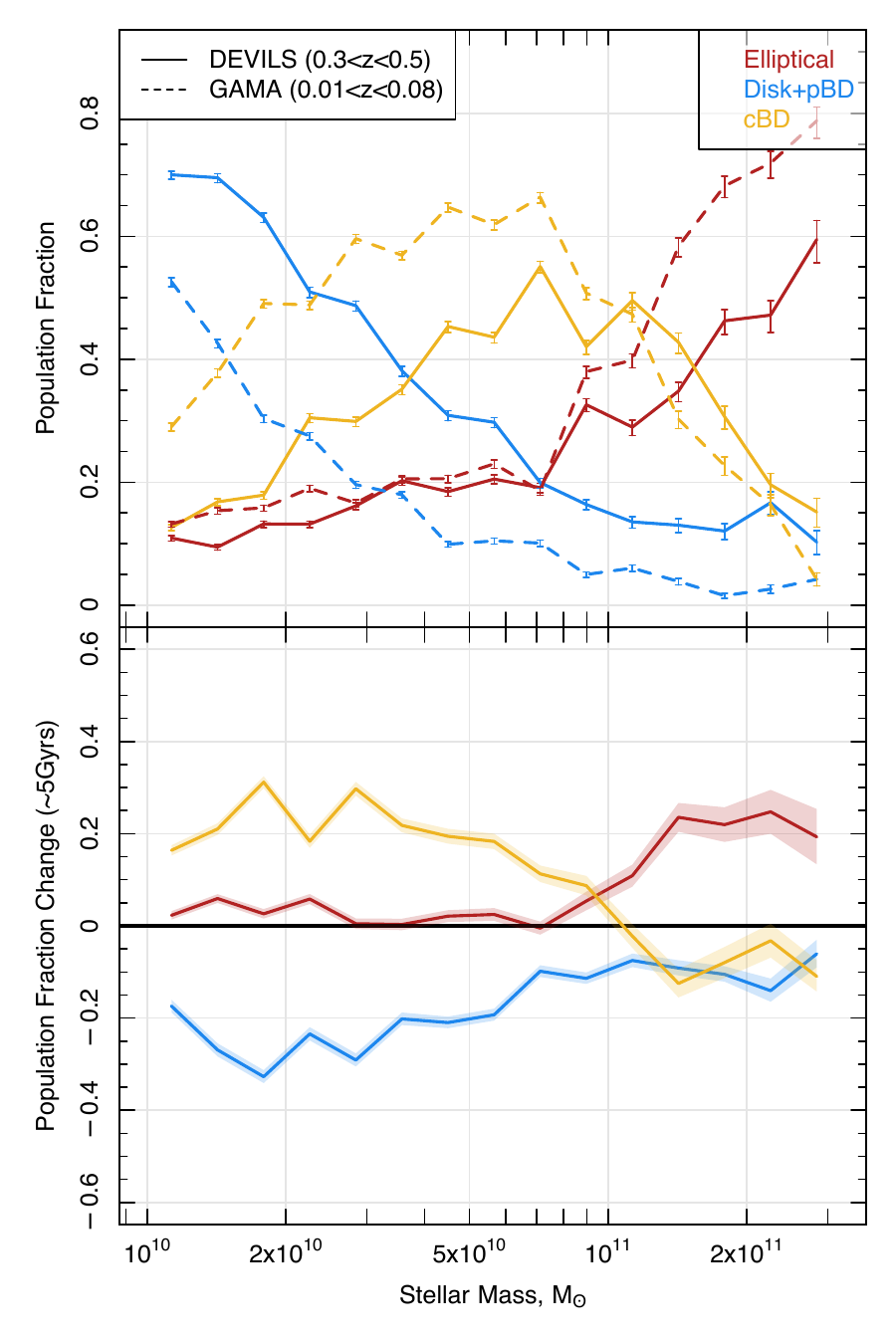}
\caption{Morphological fractions as a functions of stellar mass. Top panel shows the population fraction for DEVILS in solid lines, and GAMA in dashed lines. Error bars are calculated from the 1$\sigma$ confidence using the binomial distribution. Bottom panel shows the change in population fraction between DEVILS and GAMA ($i.e.$ GAMA-DEVILS from the top panel), where shaded regions represent the extremes from the errors in the top panel. }
\label{fig:massDense}
\end{center}
\end{figure}

\subsection{Local Galaxy Densities}
\label{sec:Sigma}

In this work we wish to explore the distribution of different visual morphological types as a function of local galaxy density. To define the local density in both GAMA and DEVILS, we first take a density-defining population of galaxies at log$_{10}$(M$_{\star}$/M$_{\odot}$)$>$10  (resulting in 3610 and 4109 galaxies in DEVILS and GAMA respectively) and then, for each galaxy, only consider sources within a spectroscopic radial velocity separation of  $\pm1000$\,km\,s$^{-1}$ \citep[$e.g.$ as in][]{Baldry06}. With this population, we then calculate the projected galaxy density as: 

\begin{equation}
\Sigma_{N}=\frac{N}{\pi d_{N}^{2}}
\end{equation}

\noindent where $d_{N}$ is the physical distance to the Nth nearest neighbour. In our analysis we calculate this for the 3rd, 4th and 5th nearest neighbour ($\Sigma_{3}$, $\Sigma_{4}$, $\Sigma_{5}$), and also for the average of $\Sigma_{4}$ and $\Sigma_{5}$ as in \cite{Baldry06}. However, we note that this choice does not significantly impact any of the results presented in this work. As such, for ease of comparison to existing works, we use the commonly quoted $\Sigma_{3}$ values for all of our figures. We also note here for completeness, that the spectroscopic observations of GAMA and DEVILS are undertaken with the same facility and using the same instrumental set-up and the redshift measurements are undertaken with the same code - reducing any possible redshift bias that may effect measurements of $\Sigma_{N}$.    

\textcolor{black}{One other potential bias in the measurement of $\Sigma_{N}$ is the fact that, any measurement of local galaxy density may be biased by galaxy peculiar velocities. Ideally, we would like to use cosmological redshifts alone to define the $n^{th}$ nearest neighbour. However, with observational redshift measurements we cannot easily disentangle cosmological redshifts and redshifts induced by peculiar velocities. This may lead to erroneously identified nearest neighbours, biasing $\Sigma_{N}$. This is somewhat mitigated by selecting only sources within a spectroscopic radial velocity separation of  $\pm1000$\,km\,s$^{-1}$ and using projected separations. However, it is still interesting to consider potential biases on our measurement of $\Sigma_{3}$ induced by galaxy peculiar velocities. In Appendix \ref{app:densityErrors} we estimate these potential errors using the \textsc{Shark} semi-analytic model \cite{Lagos24}, and show that they do not significantly impact the core result of this paper.}         

Before further discussing the distribution of $\Sigma_{3}$ for our samples, we also highlight that given the nature of both GAMA and DEVILS as volume-limited, uniformly-selected surveys, even at the high-density end of $\Sigma_{3}$, we are still predominantly probing group-scale environments. To show this, Figure \ref{fig:halo} displays the distribution $\Sigma_{3}$ values as a function of halo mass, for the subset of galaxies with which halo masses are measured in GAMA \citep{Robotham11} and DEVILS (Bravo, in prep). Here we find that, while there is a weak correlation between $\Sigma_{3}$ and halo mass, the typical halo mass at all local densities is still in the group regime. Galaxies without halo mass measurements in our samples are also likely at lower $\Sigma_{3}$ ($i.e.$ they are typically isolated centrals), which would further lower these relations.  This is important when both framing the results outlined in this work and comparing to existing results. As discussed in Section \ref{sec:Discussion}, the majority of previous studies exploring the evolution of the morphology-density relation probe targeted cluster fields. In terms of environmentally-driven galaxy evolution processes (specifically those induced by interactions/mergers) group and cluster environments can be very different. We highlight this here to indicate that care must be taken in the physical interpretation of the result presented below, and will return to this line of discussion in Section \ref{sec:Discussion}.

In the bottom row of Figure \ref{fig:hists} we then show the volume number density of different morphological types as a function of $\Sigma_{3}$ (where low values of $\Sigma_{3}$ indicate low-density environments). Here we see some interesting evolutionary trends between DEVILS and GAMA. Firstly, in terms of $\Sigma_{3}$ for the full sample (left panel), we see no significant evolution in the distribution of all galaxies. There is a potential hint of a decline in the number of low-density environments and an increase in the most over-dense environments. This is expected as the Universe evolves and larger scale structures accumulate under gravity. What is potentially most interesting here, is that we see different evolutionary trends for different morphological classes. Elliptical galaxies increase in number density in the most over-dense environments, but do not appear to decline in low-density environments. This could potentially suggest that galaxies are transitioning to ellipticals in over-dense structures via interactions/merges ($i.e.$ ellipticals are not leaving low-density environments for high-density environments, but are being newly-formed in high-density environments). disk \& pseudo-bulge+disk galaxies decrease in number-density at all values of $\Sigma_{3}$, but potentially decline to a greater degree in low-density environments. This potentially suggests that galaxies are morphologically transitioning away from pure-disk \& pseudo-bulge+disk, but via a process or processes that are not solely limited to the most over-dense environments (potentially minor mergers and/or secular evolution). Finally classical-bulge+disk galaxies appear to somewhat mirror the trend of disk \& pseudo-bulge+disk galaxies in low-density environments, increasing in number density, while retaining similar number densities (or declining slightly) in the most over-dense environments.  

\begin{figure}
\begin{center}
\includegraphics[scale=0.55]{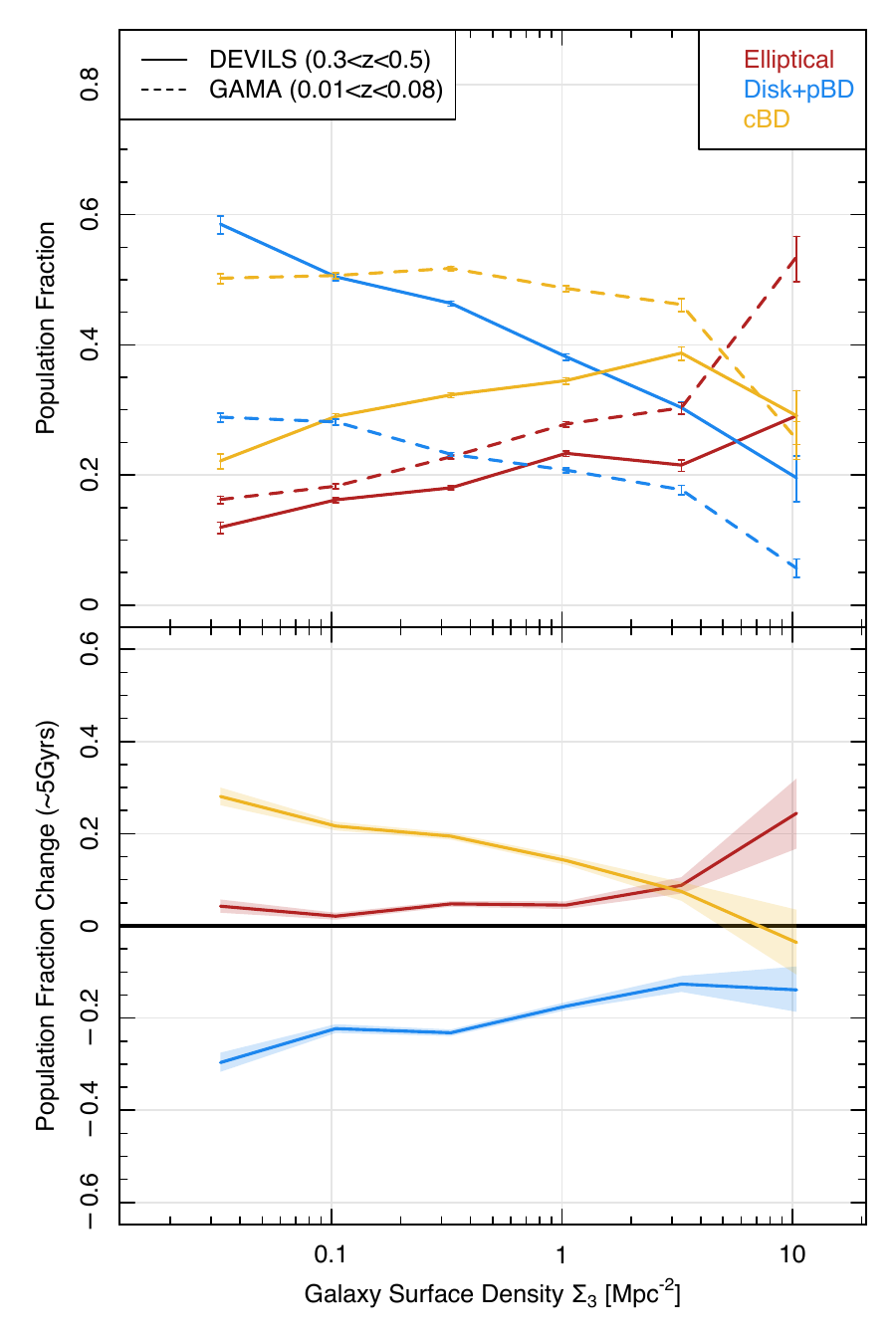}
\caption{Morphological fractions as a functions $\Sigma_{3}$. Top panel shows the population fraction for DEVILS in solid lines, and GAMA in dashed lines. Error bars are calculated from the 1$\sigma$ confidence using the binomial distribution. Bottom panel shows the change in population fraction between DEVILS and GAMA ($i.e.$ GAMA-DEVILS from the top panel), where shaded regions represent the extremes from the errors in the top panel. }
\label{fig:sigDense}
\end{center}
\end{figure}

To first order, we could tentatively suggest that these results imply a scenario where: i) in low-density environments disk \& pseudo-bulge+disk galaxies are transitioning into classical-bulge+disk galaxies through a processes which leads to the growth of classical, dispersion-supported, bulges and must remove angular momentum from disk-like pseudo-bulges \cite[potentially minor mergers that are not captured in the density-defining population. However, ][find that classical bulges have uniformly old stellar populations such that these minor mergers must not significantly affect the global ages of the bulge stellar population.]{Bellstedt24} and ii) while in higher-density environments galaxies are transitioning into ellipticals via some large-scale environmentally-regulated process (major mergers in group-scale environments?). These results will be explored further in subsequent sections, but are largely consistent with current theoretical models for morphological transitions as a function of environment.     
  
 \vspace{2mm}
  
One potential caveat here, is that we have so far only considered stellar mass distributions and local density separately. However, it is well-known that stellar mass and environment are strongly linked \citep[$e.g.$][]{Alpaslan15} - $i.e.$ there is an environment-conditional stellar mass function. Ideally, in analyses such as these one would wish to control for the same stellar mass distribution across all environments, morphologies and redshifts. However, this becomes problematic as resultant sample sizes become very small, particularly at higher redshifts. To potentially belay some concerns regarding the impact of these conditional stellar mass distributions strongly impacting our results, we calculate the population fraction as a function of stellar mass histogram in ranges of $\Sigma_{3}$ for both DEVILS and GAMA (essentially removing the normalisation of the stellar mass function). These are shown in the left and middle panels of Figure \ref{fig:MassFuncSig}, where different colours represent different local-density environments. What we wish to highlight here is that first, the shape of these distributions are largely similar across different environments (at the stellar masses used in this work). As expected, we do see some increase in the number of massive galaxies as we move from lower to higher density environments (this is consistent with previous studies that find a top-heavy stellar mass function with increasing environmental over-density). Here, to test if this variation is significant, we do produce stellar-mass-matched distributions of galaxies to align these environments, and find that, within errors and with limited sample sizes, it does not significantly change our results. However, we do not use this sample in the rest of this paper, as the stellar-mass-matching process significantly reduces the sample size available. 

While we do not use a fully environment and stellar mass-matched sample, we can also explore if these conditional stellar mass distributions would bias any evolutionary trends found in our work. As we wish to compare the evolution between DEVILS and GAMA at a fixed local density, we can compare the change in conditional stellar mass distributions between the two surveys. The right panel of Figure \ref{fig:MassFuncSig} then shows the difference between GAMA and DEVILS stellar mass distributions as a function of $\Sigma_{3}$ ($i.e.$ lines in the middle panel minus lines in the left panel) with the same colouring for local density. We find that there is no strong evolutionary change between DEVILS and GAMA in terms of the stellar mass distribution as a function of local density ($i.e.$ all lines are close to zero, with no significant trends as a function of stellar mass). This is important, as it suggests that any change we see in the distribution of morphological types, as a function of local density, is likely \textit{not} driven by a change in the stellar mass distribution of galaxies between GAMA and DEVILS. This will be revisited later.

One other caveat we note here, is that this analysis does not consider the fact that galaxies grow in stellar mass (largely via star-formation, but also mergers) over the epochs probed here ($i.e.$ progenitor bias). While we have shown above that the stellar mass distribution as a function of local density does not evolve significantly between DEVILS and GAMA, this galaxy growth could also subtly impact the density-defining population at each epoch. For example, a log$_{10}$(M$_{\star}$/M$_{\odot}$)$=$10 galaxy used to define the local over-density in GAMA, would not have been used to define the local over-density in DEVILS (as $\sim$4\,Gyr earlier it would have log$_{10}$(M$_{\star}$/M$_{\odot}$)$<$10). To simplistically attempt to estimate if this has an impact on our derived results, we explore using an evolving density-defining population based on the growth of typical galaxies. From \cite{Thorne21} we take the typical SFR of a log$_{10}$(M$_{\star}$/M$_{\odot}$)$=$10 galaxy in DEVILS from the normalisation of the star-forming sequence at $z$$\sim$0.4 as 2.5\,M$_{\odot}$\,yr$^{-1}$. Assuming all galaxies grow at this rate would imply that a log$_{10}$(M$_{\star}$/M$_{\odot}$)$=$10 galaxy in DEVILS is comparable to a log$_{10}$(M$_{\star}$/M$_{\odot}$)$=$10.3 galaxy in GAMA (assuming 4\,Gyr of evolution between the two surveys and constant star-formation). We then set our density-defining population of galaxies in GAMA to only contain systems at log$_{10}$(M$_{\star}$/M$_{\odot}$)$>$10.3 and repeat our analysis. We find that this only very slightly changes the GAMA distributions and does not impact any of the overall trends seen in our results.

\section{Morphological Fractions as a Function of Stellar Mass and Local Density}
\label{sec:Fractions}

In this section we next compare the morphological population fractions as a function of both stellar mass and local density (as traced by $\Sigma_{3}$). In all figures 1$\sigma$ confidences on each point are calculated from the binomial distribution estimated using a beta distribution following the procedure of \cite{Cameron11}. First in Figure \ref{fig:massDense}, we show the population fractions as a function of stellar mass for both DEVILS (solid lines) and GAMA (dashed lines). In both DEVILS and GAMA we see the well-known trends of galaxy morphology with stellar mass. Disk-dominated systems (blue) are far more prevalent at low stellar masses, and decline in frequency as we move to higher stellar masses. Conversely, elliptical galaxies are predominantly found at high stellar masses, and decline as we move to lower stellar mass. Disk + compact bulge (cBD) systems have low population fractions at both high and low stellar masses, and peak at around M$^{*}$ (log$_{10}$(M$_{\star}$/M$_{\odot}$)$\sim$10.6).  We note here, that we are only considering stellar masses in our sample which are complete to all morphological types, and as such, these results are not driven by sample incompleteness at the low stellar mass end. We also note that the population fractions as a function of stellar mass presented here, are consistent in terms of normalisation and shape with the results found for $z$$<$0.1 galaxies in SDSS \citep{Nair10} and in the EAGLE simulation \citep{Pfeffer23}.  
  
 While the overall distributions of the population fractions are similar between GAMA and DEVILS, we do start to see interesting evolutionary trends as a function of stellar mass. This is highlighted by the bottom panel of Figure \ref{fig:massDense}, which shows the difference between the GAMA population fraction at a given stellar mass and the DEVILS population fraction at the same stellar mass. A positive value here indicates that the population has increased in fraction over the last $\sim$5\,Gyr. Firstly, at low stellar masses we see a largely uniform decrease in the fraction of disk \& pseudo-bulge+disk galaxies at all stellar masses. This is more extreme at the low stellar mass end, where it is mirrored by an increase in classical-bulge+disk galaxies. This is potentially indicative of disk \& pseudo-bulge+disk galaxies transitioning to classical-bulge+disk galaxies at lower stellar masses.  However, at higher stellar masses (log$_{10}$(M$_{\star}$/M$_{\odot}$)$\sim$11) we see a decline in both classical-bulge+disk galaxies and disk \& pseudo-bulge+disk galaxies, and a mirrored increase in elliptical systems. At lower stellar mass ellipticals appear unchanged in population fraction between DEVILS and GAMA.

\begin{figure*}
\begin{center}
\includegraphics[scale=0.55]{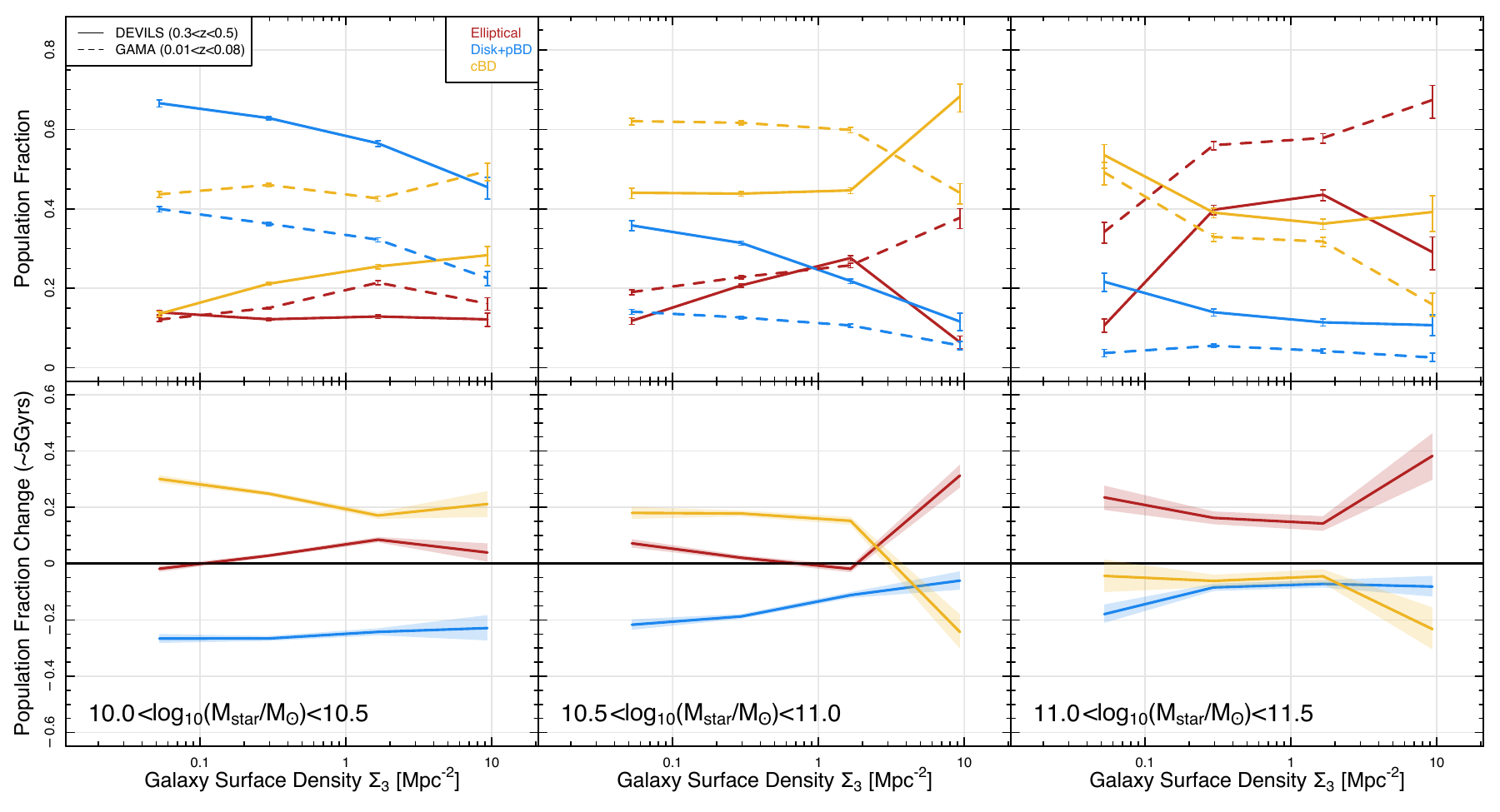}

\caption{The same as Figure \ref{fig:sigDense}, but now split into three stellar mass ranges, given in the bottom left legend. } 
\label{fig:sigDenseMasscut}
\end{center}
\end{figure*}

  \vspace{2mm}
  
 However, this result says nothing about the potential of environmental process driving these transitions.  Next in Figure \ref{fig:sigDense} we show the evolution of the population fractions now as a function of  $\Sigma_{3}$, and once again show the evolutionary change in the bottom panel. Firstly, here we see the well-known morphology-density relation for both GAMA and DEVILS samples, this is consistent with previous observational relations \citep{Dressler80, Houghton15, Siudek22} and simulation relations \citep{Pfeffer23}. For both samples, we see large fractions of disk \& pseudo-bulge+disk galaxies in low-density environments, with the fraction declining with $\Sigma_{3}$, moderate fractions of classical-bulge+disk galaxies in low-density environments with increasing fractions with $\Sigma_{3}$, until the relation turns over and drops in the most over-dense environments, and low fractions of elliptical galaxies in low-density environments, with an increasing fractions as we move to high-density environments. 

However, with the benefit of the consistently analysed and minimally-biased samples across DEVILS and GAMA, we can now also explore the time evolution of this morphology-density relation over the last $\sim$5\,Gyr. The bottom panel of Figure \ref{fig:sigDense} highlights this change. Here we see that the fraction of disk \& pseudo-bulge+disk galaxies declines considerably across all environments, but has a weak trend with $\Sigma_{3}$, suggesting less decline in higher-density environments. Once again, we see this mirrored in the classical-bulge+disk galaxies, which show a considerable increase in their fraction in low-density environments, but very small change in the most dense environments. Finally, we find an increase in the population fraction of ellipticals in high-density environments, mirrored by a decline in disk \& pseudo-bulge+disk galaxies and classical-bulge+disk galaxies. We remind the reader here that in Section \ref{sec:Sigma} we showed that there is no clear difference in the local-density selected stellar mass distributions between DEVILS and GAMA. As such, any evolutionary trend in the morphological population fractions as a function of local density, is likely not driven by a change in the stellar mass distribution within a given environment.      

Combined, these trends once again suggest that potentially in low-density environments evolutionary processes are leading to the transition of disk \& pseudo-bulge+disk galaxies into classical-bulge+disk galaxies, while in high-density environments, both classes are potentially transitioning into ellipticals. While this is consistent with previous results and the theoretical interpretations of the impact of over-dense environments on galaxy morphology, we are now witnessing this directly in well-controlled and consistently measured samples across DEVILS and GAMA - no longer inhibited by potentials biases in sample selection or methodology.

\subsection{The morphology-density relation at a fixed stellar mass}

In the above we have considered stellar mass and local-density separately. As noted in the previous section, taking fully stellar-mass matched samples as a function of local density is problematic. However, it is possible to split the samples used above into smaller stellar mass ranges, and explore if these trends remain when taking galaxies at a fixed stellar mass. In Figure \ref{fig:sigDenseMasscut} we show the same as Figure \ref{fig:sigDense}, but now split into three stellar mass ranges: 10.0$<$log$_{10}$(M$_{\star}$/M$_{\odot}$)$<$10.5 (low stellar mass),  10.5$<$log$_{10}$(M$_{\star}$/M$_{\odot}$)$<$11.0 (intermediate stellar mass) and 11.0$<$log$_{10}$(M$_{\star}$/M$_{\odot}$)$<$11.5 (high stellar mass). Note here that we reduce the number of bins in local-density values to retain adequate sample sizes in each bin. Here we see a number of interesting trends, which we will discuss for each morphological type separately: 

\subsubsection{Disk \& pseudo-bulge+disk galaxies}

First, we see that the morphology-density relation for disk-dominated galaxies holds true for all stellar mass ranges and epochs, $i.e$ the population fraction for disk-dominated galaxies declines with higher local-density at all stellar masses and in both DEVILS and GAMA (modulo the high stellar mass sample in GAMA where the numbers of sources are very small). This suggests that the observed correlation of a deficit of disk-dominated galaxies in high-density environments is not simply a consequence of environment-conditional stellar mass distributions.  We also see the overall trend of population fractions of disk-dominated galaxies declining between DEVILS and GAMA at all stellar masses. This is most strongly evident at low stellar masses, and is largely uniform across all local-densities. However, there is also a weak trend of this decline being most significant in low-density environments at all stellar masses. This potentially indicates that there is both a dominant environmentally-agnostic process morphologically transforming disk-dominated galaxies into more evolved classes (minor mergers and/or secular evolution?), and a weaker environmentally-driven process which leads to the morphology-density relation (major mergers?). In combination, this indicates that there is a decline in the population of disk-dominated galaxies at all stellar masses probed in this work over the last $\sim$5\,Gyr, and that the strength of this decline is inversely correlated with stellar mass, and to a lesser degree, inversely correlated with environment. Note, that this says nothing about the increasing prevalence of star-forming disk-dominated star-forming systems at lower stellar masses than we consider here ($i.e.$ downsizing).           
 
 \subsubsection{Elliptical Galaxies}

As with the disk-dominated galaxies, the morphology-density relation for ellipticals largely holds true across all stellar mass ranges and epochs. For GAMA we see an increasing fraction of ellipticals to higher local-density environments for all stellar masses. However, for DEVILS this picture is slightly more complex. In the lowest stellar mass galaxies we see no trend with local density. However, we note here that the sample sizes are very small. At intermediate and high stellar masses, we see an increasing elliptical population fraction from low- to high-density, which then decreases to the highest-density bin. We note here, that this is mirrored by an increase in the classical bulge+disk population. Interestingly, this is somewhat opposite to the trends seen in GAMA at the same stellar mass. 

One potential explanation for this, is the preferential misclassification of ellipticals as classical bulge+disk systems in DEVILS high density environments. While this is somewhat unlikely to only affect galaxies in high-density environments, potentially the crowded nature of these fields leads to misclassification. For example, tidal interactions may cause the outer parts of elliptical galaxies to look more disturbed in higher density regions, making them appear more bulge+disk-like. To explore this further, we first take all 0.3$<$$z$$<$0.5 and 10.5$<$log$_{10}$(M$_{\star}$/M$_{\odot}$)$<$11.5 classical bulge+disk galaxies with $\Sigma _{3}$$>$1 in DEVILS  (resulting in 92 sources) and compare the spread of visual classifications across all classifiers. We find that only 9 of these sources have even one classifier visually identify them as an elliptical ($i.e.$ there is strong agreement from classifiers that these are in fact classical bulge+disk galaxies). Even if we assign all of these objects as ellipticals and reproduce Figure \ref{fig:sigDenseMasscut}, it does not change any of the observed  trends. We then also re-visually inspect the 92 sources selected above and confirm that all sources are consistent with a classical bulge+disk classification. Thus, we do not believe this trend is a consequence of visual misclassification. This then suggests that in the highest-density environments we do see a strong evolution of the elliptical population between DEVILS and GAMA at a fixed stellar mass, with a large increase in the elliptical population fraction over the last 5\,Gyr. This increase is most extreme in intermediate stellar mass galaxies, where there is little-to-no evolution in low-density environments. We do see more uniform evolution in the highest stellar mass galaxies, once again suggesting that the morphological transition to ellipticals is strongly correlated with stellar mass, but there is also a correlation with local density (as for the disk dominated systems), with the strongest evolution in the highest local density bin.            
  
  \subsubsection{Classical bulge+disk galaxies}    
 
Finally, considering the classical bulge+disk galaxies we see arguably the most interesting trends. At low stellar masses we find a weakly increasing population fraction as a function of local density in both DEVILS and GAMA, suggesting that classical bulge+disk galaxies are more prevalent in higher-density environments. We also find that the decline in disk \& pseudo-bulge+disk galaxies is almost identically mirrored in the increase of classical bulge+disk galaxies at all local densities, and for both DEVILS and GAMA. This adds weight to the argument that at these stellar masses, some environmentally-agnostic process is driving the transition from disk-dominated to classical bulge+disk galaxies, while a second environmentally-driven process is also converting disk-dominated galaxies into classical bulge+disk galaxies, leading to the low mass morphology-density relations observed here.   

 At intermediate stellar masses we see a flat population fraction with increasing local density for both GAMA and DEVILS, until we reach the highest density environments. Interestingly, the population fractions show opposite trends for GAMA and DEVILS, where the population fraction increases for DEVILS and decreases for GAMA (mirroring the elliptical population). This suggests that in the highest-density environments, there is potentially a strong evolution of classical bulge+disk galaxies into ellipticals at this stellar mass range. 

At the highest stellar mass we actually see an anti-correlation between classical bulge+disk population fractions and local-density in both GAMA and DEVILS, showing that these systems are less common in high-density environments. This potentially indicates that these galaxies have been morphologically transformed by some environmentally-driven process. We also again see a strong evolution between DEVILS and GAMA for the highest-density environments, suggesting the transition of classical bulge+disk galaxies into elliptical.          

Interestingly, these results also highlight that the observed morphology-density relation for classical bulge+disk galaxies, actually varies significantly as a function of stellar mass (both for DEVILS and GAMA). At low stellar masses we find a larger fraction of classical bulge+disk galaxies with increasing local density, at intermediate stellar masses this flattens, and at high stellar masses the relation flips to show a smaller fraction of classical bulge+disk galaxies with increasing local density. This potentially shows an evolutionary process that is correlated with both environment and stellar mass.     

\vspace{2mm}

In summary, when exploring samples at a fixed stellar mass the observational trends are consistent with our previous assertions that: \\

\noindent $\bullet$ At all stellar masses we see a potential transition from disk-dominated galaxies to classical bulge+disk galaxies via a process likely un-correlated with environment which occurs at all local-densities. This leads to an overall decline in disk-dominated galaxies and increase in classical bulge+disk galaxies. This process is also strongest at low stellar masses. Potentially, there is also a weaker environmentally-driven process, which is transitioning disk-dominated galaxies to classical bulge+disk galaxies in high-density environments. This results in a classical bulge+disk fraction with is correlated with local density and a disk-dominated fraction which is anti-correlated with local density.  However, care must be taken here, as this classical bulge+disk population are traditionally the most difficult to robustly visually classify. In addition, the spectro-structural decomposition work of \cite{Bellstedt24} at $z$$\sim$0 find that the bulges in classical bulge+disk galaxies have very old and uniform age stellar populations, suggesting that they have had little accretion of younger stellar material over their universal history (as would occur in mergers). This would argue against the fact classical bulges are being formed in the local Universe. However, our current data does show a large increase in the number density of classical bulge+disk galaxies over the epochs probed here. With the limits of our current analysis we can say little more.\\ 

\noindent $\bullet$ At intermediate stellar masses, we see increase in the elliptical population fraction, but only in the highest-density environments. This is  mirrored by a decline in the classical bulge+disk population, suggesting that some aspect of the local-density is driving a visual morphological change from classical bulge+disks to ellipticals.  \\ 

\noindent $\bullet$ At high stellar masses, we see increase in the elliptical population fraction at all local densities, but the most significant increase in the highest-density environments, which is mirrored by a decline in the classical bulge+disk population. This suggests that at high stellar masses classical bulge+disk galaxies are transitioning into ellipticals, and that this process is accelerated by the local environment. We also see an anti-correlation between classical bulge+disk population fractions and local-density, potentially highlighting that high stellar mass classical bulge+disk galaxies have been reduced in over-dense environments via some environmentally-driven process (consistent with the previous statement). \\

\begin{figure*}
\begin{center}
\includegraphics[scale=0.8]{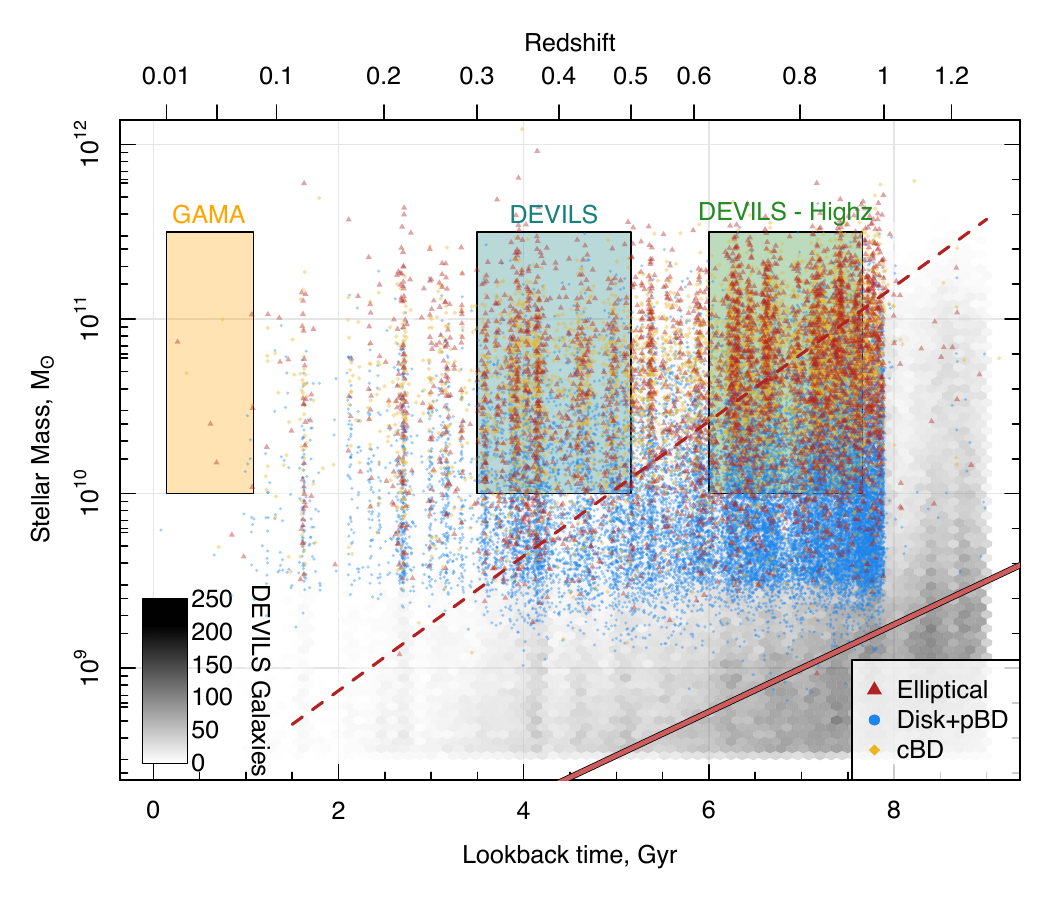}

\caption{Similar to the panels in Figure \ref{fig:sample}, but now showing the selection of a high-z sample in DEVILS. Note for clarity, we now show the x-axis in look back time. The grey shaded regions show the full DEVILS-D10 sample including spectroscopic, grism and photometric redshifts. The coloured points show sources with morphological classifications in \citet{Hashemizadeh21}. The red dashed line shows the spectroscopic completeness limit from Figure \ref{fig:sample}, while the solid light red line shows the full spectroscopic+grism+photometric completeness line from \citet{Thorne21}. The orange and teal polygons show the regions already used in this work which are spectroscopically complete. The green polygon shows the new high-z sample, which is morphologically complete, but relies on the use of grism and photometric redshifts.} 
\label{fig:sampleHighz}
\end{center}
\end{figure*}

\section{Extending to Higher Redshifts}
\label{sec:highz}

In the previous section we calculated the evolution of the morphology-density relation using a highly-robust sample of spectroscopically-confirmed galaxies, with similar rest-frame imaging for morphological classifications. However, within the DEVILS sample, it is also possible to extend this further to earlier look back times, including objects with, lower precision, grism and photometric redshifts \citep[$e.g.$ see][for a similar approach applied to galaxy merger rates]{Fuentealba25}. This will extend the evolutionary baseline with which to probe the evolution of the morphology density relation.  

However, before proceeding there are a number of caveats to such an analysis. Firstly, the use of lower accuracy and precision photometric redshifts may bias measurements of $\Sigma_{3}$ due to chance alignments of sources and/or catastrophic redshift errors. We aim to quantify this error in the following section, but note that it may still bias the results. Secondly, the use of the HST-ACS F814W filter for morphological classifications in DEVILS, means that rest-frame imaging at higher redshifts is shifter to bluer rest-frame wavelengths. For example, at $z$$\sim$0.8 the F814W filter probes the rest-frame at $\sim$4500\AA\ ($i.e.$ $g$-band) instead of $\sim$5800\AA\ ($i.e.$ $r$-band) at $z$$\sim$0.4. Galaxies are known to show different structures/morphological characteristics as a function of rest-frame wavelength \citep{Kelvin12}. As such, this could bias morphological classifications towards bluer ($i.e.$ more star-forming) components - potentially leading to more weight being placed on disk structures over bulges. However, this difference is somewhat small, and observations remain above the rest-frame 4000\AA\ break for all sources in the sample when deriving morphological classifications, and as such likely trace similar structures. Note, that spatial resolution comparison in morphological classifications between epochs should not be an issue here, as the HST-ACS imaging data still has a physical spatial pixel scale of $\sim$0.38\,kpc at $z$=0.8.      

With these caveats in place, we next define a higher redshift sample within DEVILS-D10 with which to compare to our existing results. This selection is shown in Figure \ref{fig:sampleHighz}. Here we now show the full DEVILS-D10 sample including sources with only grism or photometric redshifts. The grey-shaded background shows the full DEVILS-D10 sample, while the coloured points show the sources with morphological classifications in \cite{Hashemizadeh21}. Redshifts here come from the DEVILS-D10 master redshift catalogue which contains spectroscopic, grism and photometric redshifts for all sources in the D10 region (this will be presented in the DEVILS DR1 data release paper, Davies et al, in prep). We select a high-z region which covers the same $\Delta$lookback time as the DEVILS sample already used in this work ($\sim$1.6\,Gyr)  and the same stellar mass range as previous samples. This covers 0.625$\lesssim$$z$$\lesssim$0.924. From Figure \ref{fig:sampleHighz} we see that this sample is complete in terms of stellar mass \citep[lying far above the light red stellar mass completeness line of][]{Thorne21} and morphological classifications, but not in spectroscopic redshifts. For galaxies in this selection window, we find that $\sim48$\% have a secure spectroscopic redshift, $\sim8$\% have a secure grism redshift: 0.7\% from 3D-HST \citep{Momcheva16}  and 7.3\% from the PRIsm MUlti-object Survey \citep[PRIMUS][]{Coil11}, and $\sim44$\% have a photometric redshift: 24.5\% from the COSMOS2015 catalogue \citep{Laigle16} and 20.5\% from the Physics of the Accelerating Universe Survey \citep[PAUS][]{Alarcon21} -  hereafter the DEVILS-highz sample.

\begin{figure*}
\begin{center}
\includegraphics[scale=0.6]{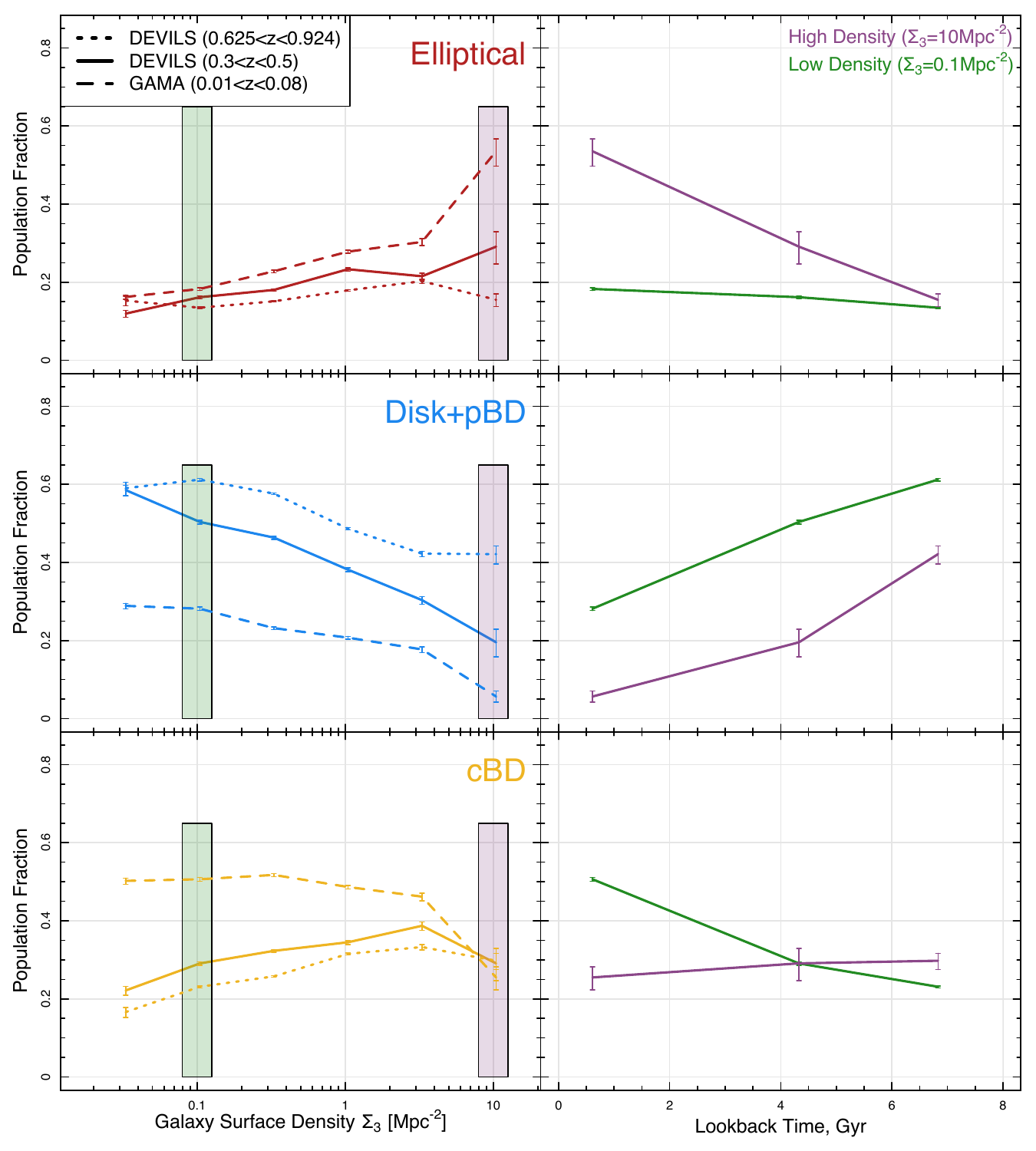}

\caption{Time evolution of morphological population fractions now including the DEVILS-Highz sample. The left columns shows population fraction as a function of $\Sigma_{3}$ for each morphological class at the three epochs used in this work. $\sigma$ confidences on each point are calculated from a beta distribution following the procedure of \citet{Cameron11}. In the right column we show the time evolution of the population fractions for high-density (purple) and low-density ( green) regions. These are also indicated in the left column as coloured rectangles. } 
\label{fig:highz}
\end{center}
\end{figure*}

Using this sample, we then repeat the measurement of projected number densities, $\Sigma_{3}$, following the methodology outlined in Section \ref{sec:Sigma} and calculate the morphological population fractions as in Section \ref{sec:Fractions}. This is shown in Figure \ref{fig:highz}. As noted previously, the use of imprecise photometric redshifts in the definition of $\Sigma_{3}$ could potentially lead to errors/biases. To quantify this, when calculating $\Sigma_{3}$ for the DEVILS-highz sample, we repeat the analysis for 1000 realisations, where in each realisation we randomly sample the galaxy photometric redshifts within the photometric redshift errors in the DEVILS-D10 catalogue, recalculate $\Sigma_{3}$ and then measure the population fractions for each morphological type. We describe this process in more detail in Appendix \ref{app:photoz} and replicate Figure \ref{fig:highz} using the median population fractions from these realisations. In summary, this does not significantly change the results presented in this work.                     

\begin{figure*}
\begin{center}
\includegraphics[scale=0.6]{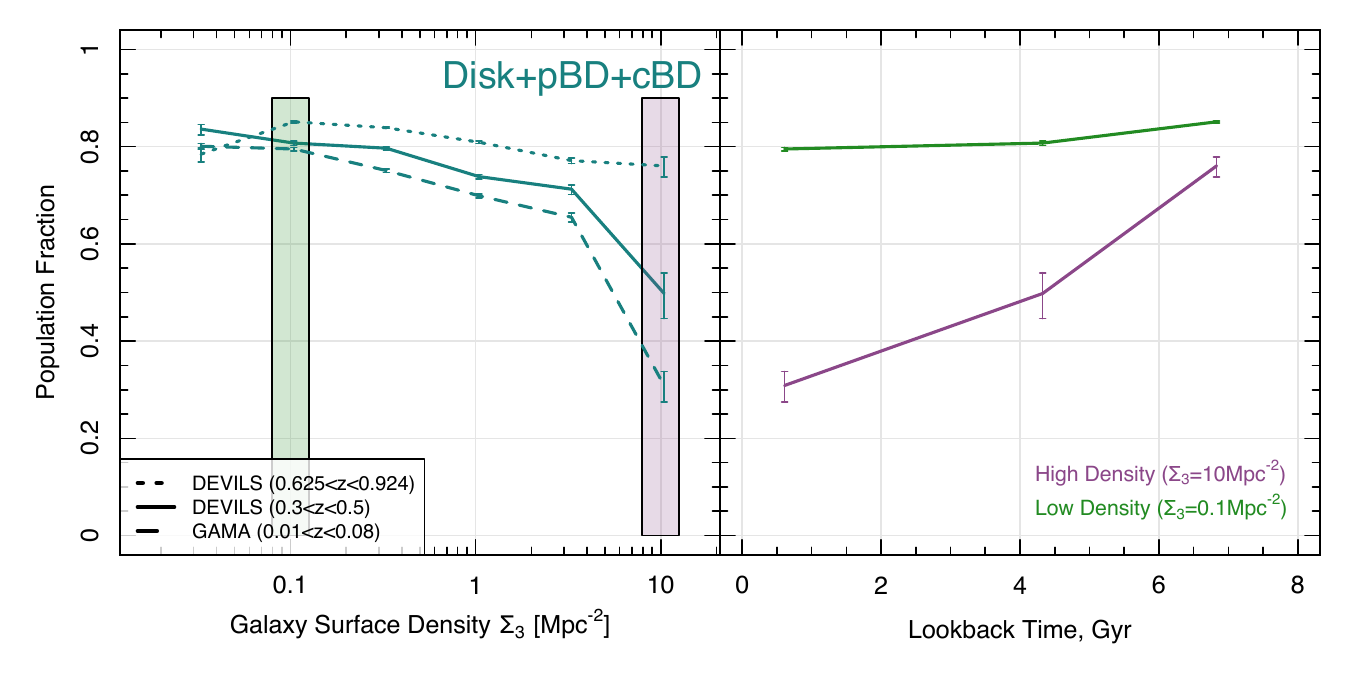}

\caption{The same as \ref{fig:highz} but showing the time evolution of population fractions for all galaxies of any type containing a disk (D + pBD + cBD).} 
\label{fig:allDisk}
\end{center}
\end{figure*}

In the left column of Figure \ref{fig:highz} we show the population fractions for each morphological type now at the three epoch probed in this work. We see clear evolution in the population fractions, with the new DEVILS-highz sample being consistent with the evolution between DEVILS and GAMA derived earlier in this work. Firstly, this adds weight to the argument that the potential biases noted above are likely not strongly driving the results seen here. Across all environmental densities, we see that the fraction of elliptical galaxies increases with time, disk+pBD galaxies decrease with time, and cBD galaxies largely increases with time. However, we do also see different evolution of the morphological populations as a function of $\Sigma_{3}$. To highlight this, the right column of Figure \ref{fig:highz} shows the time evolution of the population fractions in both high-density ($\Sigma_{3}$=$10$\,Mpc$^{-3}$) and low-density ($\Sigma_{3}$=$0.1$\,Mpc$^{-3}$) environments. We see consistent results with the previous section, but now extended in evolutionary baseline. We find that, for 10$<$log$_{10}$(M$_{\star}$/M$_{\odot}$)$<$11.5 galaxies, over the last $\sim$7\,Gyr ellipticals have increased in population fraction by $\sim0.4$ in high-density environments, but remain relatively consistent in low density environments.  Disk+pBD galaxies decrease in population fraction by $\sim0.3$ somewhat uniformly at all local densities, while cBD galaxies increase in population fraction by $\sim0.3$ in low-density environments, but show very little change in high-density environments.  Once again, these results can be interpreted as first, over the last $\sim$7\,Gyr a process, which is not strongly correlated with high-density environments, is morphologically transforming disk+pBD galaxies into cBD galaxies. While in high-density environments a process is morphologically forming elliptical galaxies. Finally, to highlight this further, in Figure \ref{fig:allDisk} we show the same as Figure \ref{fig:highz} but now combining all galaxies of any type containing a disk (D + pBD + cBD). As expected, with find very little change in population in low density environments and a large decline in the population in high density environments - consistent with an overall morphological change from these disk systems to ellipticals. Potential astrophysical processes which may be the root cause of these transitions are discussed in the following section.

\section{Comparison to Literature}
\label{sec:Discussion}

In the previous sections, we highlighted the evolutionary trends of morphological passive fractions as a function of local galaxy density. In this section, we compare to existing results for the evolution of the morphology-density relation and discuss physical interpretations to potentially explain the trends seen in our results. However, as discussed previously, comparing the time evolution of both morphological classes and environment (and doubly the combination of both) is problematic, due to differences between sample selections ($.i.e.$ redshift and stellar mass ranges, general population vs clusters fields), data ($i.e.$ imaging resolution and bands used, spectroscopic vs photometric redshifts) and methodologies (morphological classification approach, tracer of environment, choice of $N$ in $\Sigma_{N}$ estimation), etc \cite[see][for discussion around morphological differences in samples selected based on local density vs cluster-centric distance]{Vulcani23}. Differences in these can potentially lead to quite significant biases and/or lead to results that cannot easily be compared. As such, we largely attempt not to quantitatively compare to existing results but to qualitatively compare to evolutionary trends. Our approach here is to provide a self-consistent measurement of the evolution of the morphology-density relation between DEVILS and GAMA, largely avoiding these comparison issues, and not relying on comparisons to existing works. That said, overall we do find strong qualitative agreement with existing works in this area. We also note, that the majority of existing works specifically target rich cluster environments, while we probe the typical galaxy population in un-targeted fields. In addition, given the volume limitations on both GAMA and DEVILS, even in our high-density environments we are typically probing group-like structures and not massive rich clusters. This is important to consider when comparing to previous works. 

First, briefly considering only the local Universe, early studies into the morphology-density relation found a clear separation between morphological types as a function of local density, consistent with our results \citep[$e.g.$][]{Dressler80, Postman84}.  These results have subsequently been updated finding similar results \citep[$e.g.$][]{Houghton15}, and show the same morphology-density trends that we see here. At higher redshifts \cite{Poggianti08} explore spectroscopically-confirmed cluster members at 0.4$<$$z$$<$0.8 in the ESO Distant Cluster Survey (EDisCS), and find the clear local galaxy density separation between ellipticals, spirals and irregulars. This is also found using colour-selected galaxies in \cite{Siudek22}.  While, at higher redshifts still, \cite{Sazonova20} explore the morphology density relation in four rich clusters at $1.2<z<1.8$ and find that there they have an over-abundance of bulge-dominated systems in comparison to the field. This suggests that in the most over-dense regions the morphology-density relation is already in place at this epoch \citep[also see ][]{Mei23}. We highlight this here to note that the morphology-density relation is observed  over the last $\sim$10\,Gyrs of Universal history, and the trends for morphological population fractions derived in this work are consistent with this existing body of work. 

The first real approaches to explore the \textit{evolution} of the morphology-density relation came with \cite{Dressler97}, who mapped the morphology-density relation in 10 rich $\sim0.5$ clusters using HST. They then compared the evolution of population fractions in comparison to the $z$$\sim$0 clusters in \cite{Dressler80}, finding a decrease in the elliptical galaxy population with time, but a strong increase in the lenticular fraction. They suggest different morphological population fractions for, essentially, relaxed and unrelaxed clusters, and argue that this is evidence that elliptical formation is actually driven in the group-phase, which predates rich cluster formation. Given in this work, we primarily probe group-like environments, our results do not contradict this hypothesis. We see that in group-like local densities the fraction of ellipticals is increasing with time. As noted previously, given the volume limitations of both GAMA and DEVILS we do not probe the rich cluster environments similar to those mapped by \cite{Dressler97}, so cannot comment further.              

Moving to higher redshifts, \cite{Smith05} studied a sample of $z\sim1$ clusters over a range of local densities, with morphologies derived from HST imaging and largely photometric redshifts. They then directly compare this to $z\sim0$ samples, and suggest that in high-density environment, the elliptical+lenticular fraction increases by $\sim0.2$ since $z\sim1$, while in low-density environments the elliptical+lenticular fraction stays the same - consistent with the results we present here. \cite{Postman05} expanded on this to also morphologically classify galaxies using HST-ACS data over a range of epochs, and define local galaxy density based 7th nearest-neighbour, $\Sigma_{7}$. They use a collection of both spectroscopic and photometric redshifts, and study the morphology-density relation for cluster fields at both $z\sim0.4$ and $z>0.8$. They first find that the morphology-density relation is in place by $z\sim1$ and consistent population fractions to \cite{Smith05} . They also highlight population fractions of ellipticals are largely similar to lower redshift samples from $e.g.$ \cite{Dressler80} and \cite{Postman84},  similarly to \cite{Dressler97}, but they also note that there appears to be strong evolution in lenticular ($\sim$cBD) and spiral ($\sim$disk+pBD) galaxies - with lenticulars increasing and disks declining (also consistent with our work). Finally, similar to \cite{Smith05}, they also show that if they combine lenticulars and elliptical into a single class the population fraction increases by $\sim0.3$ over the last $\sim8$\,Gyr in higher-density environments, largely consistent with our results for non-cluster fields.

\textcolor{black}{Following this, \citet{vanDerWel07} explored the evolution of the field and cluster morphology density relation from $z\sim1$ to the present day. They find that, for stellar mass-selected samples, the fraction of ellipticals+lenticular galaxies in cluster-like environment does not change significantly over the epoch probed, and thus the morphology density relation does not evolve significantly since $z\sim1$. This first appears somewhat at odds with our current results. However, we note that \citet{vanDerWel07} is considering cluster-like environments (which we do not probe here) and their morphological `early-type' classification would include many of the cBD galaxies in our sample - where we do not see significant evolution in over-dense environments. For example, in our Figure \ref{fig:sigDenseMasscut} for galaxies at log$_{10}$(M$_{\star}$/M$_{\odot}$)$>$10.5 (comparable to the van Der Wel sample), we see an increasing elliptical and declining cBD population in high density environments. If combined, these classes would show little evolution. }
 
 \textcolor{black}{Next, \citet{Tasca09} also explored the evolution of the morphology-density relation from $z\sim1$, using samples selected from zCOSMOS, and find that the segregation of morphological types with environment is already in place by $z\sim1$. They find evolution of the in morphology density relation with, in over-dense environments, the fraction of early type galaxies inceasing with time and late-type galaxies declining with time - consistent with our results. They also suggest that while the morphology density relation is in place at $z\sim1$ it is flatter than in the local universe - also consistent with our results. Finally, they argue that the evolution of the morphology-density relation is strongly dependant on stellar mass, with lower stellar mass galaxies being more strongly impacted by their environment. This is also found when using group environments directly in zCOSMOS \citep{Kovac10}, and is broadly in agreement with our findings here ($i.e.$ Figure \ref{fig:sigDenseMasscut}).     }             

Most recently the Euclid Collaboration Quick Data Release select early type galaxies based on S\'{e}rsic Index and $u-r$ colour, based on state-of-the-art Euclid imaging. They then define the local galaxy density using photometric redshifts and also 7th nearest-neighbour densities \citep{Cleland25}. Noting here that they only consider relatively high density environments ($\Sigma7>10$), over similar stellar mass ranges to our sample, they find that the fraction of early type galaxies increases with $\Sigma_{7}$ and that the fraction of early type galaxies is increasing with time, across all $\Sigma_{7}$ values. To first order, assuming all early-type galaxies selected on S\'{e}rsic Index and colour are ellipticals, this is consistent with our results.      

Finally, it is also worth noting here, that the morphology density relation and its evolution is also observed in hydrodynamical simulations, $e.g.$ \cite{Pfeffer23}, who find that that the $z\sim0$ morphology-density relation derived for the EAGLE simulation is very similar to that observed in, $e.g.$ \cite{Houghton15}. However, using the EAGLE simulation, \cite{Pfeffer23} go further and identify three processes that they suggest result in the observed morphology-density relation: i) the transformation of disk-dominated to lenticular systems through gas stripping in high density environments, ii) formation of lenticulars through mergers and AGN feedback in low-density environments, iii) elliptical galaxies are more prevalent in high-density simply because both morphology and density are correlated with stellar mass \citep[see also][]{Alpaslan15}. The third statement is also consistent with the inferences from \cite{vanderWel10}, who explore the morphology-density relation using SDSS, but use quiescence as a proxy for early-type morphology and dark matter halo mass as proxy for local density \citep[however, see][ for differences in these approaches]{Vulcani23}. They also argue that elliptical galaxies are more prevalent in high-density environments simply because more massive galaxies are more prevalent in high-density environments, and more massive galaxies are likely to be elliptical. $i.e$ that the environment plays no role in the transition of galaxies to elliptical-like morphologies \citep[also see][]{DePropris16}.   

The first two of these statements in \cite{Pfeffer23} are consistent with our results, where we see the likely, somewhat environment agnostic, transition of disk+pBD galaxies to cBD  galaxies at all environments. However, the third is potentially at odds with our current observations. To highlight this, we remind the reader that we see little overall change in the stellar mass and $\Sigma_{3}$ distributions between DEVILS and GAMA (left panels Figure \ref{fig:hists}) and very little difference in the distribution of stellar masses as a function of local density (Figure \ref{fig:MassFuncSig}). This largely suggests that the overall distribution of stellar masses as a function of local density does not change over these epochs, and suggests that any change in the morphological population fractions is likely not due to the evolution in the stellar mass function.  Given that we do see a strong evolution in the fraction of elliptical galaxies in higher-density environments (Figure \ref{fig:highz}), these results are potentially at odds with the arguments in \cite{Pfeffer23} and \cite{vanderWel10}.  If the large fraction of ellipticals in high-density environments is simply driven by the fact that more massive galaxies are found in high density environments, our lack of evolution in the stellar mass distribution between DEVILS and GAMA would suggest we should see no change in elliptical fraction - counter to our results. We also find that when splitting the sample into three stellar mass ranges (Figure \ref{fig:sigDenseMasscut}), that the morphology-density relation still remains, and we still see environmentally-dependant evolution between DEVILS and GAMA.             

One important point to note here, is that the simulation work of \cite{Pfeffer23} extends to higher local galaxy densities than those probed here (cluster-like), and also the work of \cite{vanderWel10} makes this statement with regards to massive group and cluster-like environments, while here our high-density environments are probing lower mass groups ($i.e.$ see Figure \ref{fig:halo}). Potentially the above statements from \cite{Pfeffer23} and \cite{vanderWel10} only hold true for the most massive, cluster-like, over-densities, which we do not probe here. This is also somewhat aligned with the statements made in \cite{Dressler97}, who suggest that the majority of the elliptical formation occurs in the group-phase. Are the trends we see in our work, simply the buildup of elliptical galaxies in group-scale environments, and would this evolutionary trend potentially disappear if we were to extend our analysis to higher density cluster-like region? Considering that elliptical galaxies likely form from major mergers, $e.g.$ \cite{Barnes91, Hopkins06, Hopkins08} although see \cite{Bournaud07}, this is potentially unsurprising. In the most massive clusters velocity dispersions are high, meaning that while galaxy interactions (stripping, tidal interactions, harassment) are high, actually major mergers are surprisingly likely less frequent \citep[$e.g.$see][]{Pearson24}. However, in groups the galaxy volume density is still relatively high, but velocity dispersions are much lower, potentially leading to higher numbers of major mergers - and increased elliptical formation. So, as our study is potentially probing groups-scale environments at the high density end, it is likely unsurprising that we may be seeing the build-up of the elliptical population in these environments that then may cease to occur if we moved to higher-density, rich cluster-like, environments. 

In summary, the morphology-density relation and evolutionary trends in population fraction we see in our data are largely consistent with existing observational data and theoretical interpretations (modulo differences in data, methodology, etc). We suggest that our results are consistent with a model where at all environments disk-dominated systems are morphologically transitioning into classical bulge+disk systems, this is potentially due to minor mergers resulting in classical bulge growth. This results in an overall decline in the disk-dominated population and rise in the cBD population. In higher density environments (groups to low-mass clusters), we see an increasing fraction of elliptical systems potentially due to major mergers of all other classes. Based on the works of $e.g.$ \cite{vanderWel10} and \cite{Pfeffer23}, this growth of ellipticals may stop in the most over-dense cluster-like environments where velocity dispersions are too high for significant number of major mergers in the satellite populations. 
 
Finally, we also reiterate here the potential discrepancies between visual-morphological classifications and those based on galaxy dynamics \cite[$e.g.$][]{Cappellari11, Guo20, FraserMcKelvie22, Rigamonti24}. These studies find that only a fraction ($\sim$50\%) of the local elliptical population are true dispersion-supported systems (slow rotators), with the remaining systems showing some rotational component (fast rotators) - potential 2-component systems with faded disks. While the evolution of this fraction is currently unknown, we must caveat that the evolutionary trends for ellipticals found in our work, could be driven by the increase in these rotation-supported but visually-defined ellipticals over time. One could argue a scenario where the formation of these faded-disk systems is increased in over-dense environments where satellite galaxies are starved of gas leading to quenching \citep[$e.g.$ see][]{Davies19a, Davies25c} and then disk fading \citep[$e.g.$][]{Bremer18}. This would also result in an increased number of visually identified ellipticals in over-dense environments, but with galaxies undergoing little true morphological change.  This is somewhat in agreement with works exploring the prevalence of slow rotators as a function of environment \citep[$e.g.$][]{Brough17, Veale17,Greene18}, who find only weak-to-no correlation between the fraction of slow rotators and environmental over-density - once controlled for stellar mass. This would suggest that environment plays little role in the formation of dispersion-supported systems. Simulations also offer some insight into this picture. \cite{Lagos18} explore the prevalence of slow rotators in the EAGLE \citep{Schaye15} and HYDRANGEA \citep{Bahe17} hydrodynamical simulations.  They find a strong correlation between incidence of slow rotators and stellar mass, with slow rotators being almost exclusively found at high stellar masses, but only a weak correlation with environment. However, they do suggest a dichotomy between central and satellite galaxies in over-dense environments, where at a fixed stellar mass, centrals do show an increasing slow rotator fraction as a function of halo mass, suggesting that for centrals at least, morphological transformations are accelerated in over-dense environments. However, this trend is not observed in satellites - potentially indicating they undergo the quenching/disk fading pathway but with little morphological change.  Given that we do not have dynamically-defined morphological classifications for this sample, and know little about evolution of fast-rotating visually-elliptical systems, we can say little more here. Thus, we end by stating that we do see evidence of an increase in visually-morphologically selected ellipticals in high-density environments, but that the physical interpretation of these trends is currently far from clear.

\section{Conclusions and Future Directions}

Following these results it is interesting to consider how we might improve this picture going forward. There are a number of key avenues we could explore, and upcoming surveys will likely provide transformative results in this area. First, it is interesting to further consider how these morphology-density trends vary with other galaxy and environmental properties, such as star-formation rate, existence of an AGN, central/satellite status, halo mass and halo phase-space location, building on existing works in this area. Projects are ongoing with DEVILS to explore these, using the wealth of galaxy and environmental properties already defined by the DEVILS team. However, the sample size and volumes probed with DEVILS is still relatively small - leading to low number statistics when subdividing the population on these properties. To make significant progress in this area requires significantly larger volumes covering a range of epochs with both high completeness spectroscopy and high resolution imaging. Fortunately, the 4MOST-WAVES \citep{Driver19} survey will provided the redshifts and environmental metrics to do just this. The WAVES-wide program will provide a local benchmark comparable to GAMA but extending $\sim2\times$ fainter and covering $\sim4\times$ the area, while the WAVES-deep program will target similar populations to DEVILS but covering $\sim15\times$ the area. Importantly, the WAVES regions will also be observed by Euclid, providing high resolution imaging with which to define galaxy morphologies. Combined, these samples will allow us to both probe a more broad range of local densities and significantly improve the number statistic on all of the results presented here.  

In a different but complementary approach, we can also look to develop more quantitative measurements of galaxy structure using techniques such as spectro-structural decompositions \cite[$e.g.$ see ][]{Robotham22, Bellstedt24}, and correlate with environment to explore the co-evolution of galaxy components and their star-formation histories, and environment. So far these studies have been limited to the very local Universe \citep{Bellstedt24} due to the lack of multi-wavelength high-resolution imaging. However, with rapid increase in area coverage of space-based imaging from HST, JWST and Euclid (and ultimately the Nancy Grace Roman Space Telescope), we will soon be able to apply these techniques to much earlier times - exploring the co-evolution of galaxy components and environment, to witness the formation of the morphology-density relation in situ. Work is currently ongoing in this area using DEVILS+HST+JWST, and in the future will be expanded using WAVES+Euclid. This will hopefully break some of the potential inconsistencies between the increase in the fraction of disk + classical bulge systems observed here, and the ages of classical bulges seen in \cite{Bellstedt24}.       

Finally, we also note that future observations exploring galaxy dynamics outside of the local Universe and their correlation with environment ($e.g.$ with MUSE or WST), will be essential in identifying the astrophysical process which drive morphological change.       
        
\vspace{5mm}

In summary, here we have explored the evolution of the morphology-density relation using robust and well-defined samples with consistent morphological classifications and environmental metrics. The aim of this study is to remove any potential biases induced by methodology/technique in exploring the time evolution of the impact of environment in the morphological evolution of galaxies. We find that: \\

\noindent $\bullet$ The distribution of morphological types changes significantly over the last $\sim$7\,Gyrs with a significant decline in pure disk \& disk + diffuse/pseudo bulge galaxies in all environments, and an increase in both disk + compact/classical bulge and elliptical galaxies at all local densities. We propose this suggests a largely environment-agnostic process is leading to the transition away from disk-dominated morphologies (potentially minor mergers and/or secular processes), but $c.f.$, the old ages of classical bulges in \cite{Bellstedt24}. \\

\noindent $\bullet$  The traditional morphology-density relation is observed at all epochs probed, and clear environmental trends with morphology remain when exploring samples at a fixed stellar mass. \\

\noindent $\bullet$  At low stellar masses, we find negative correlation between disk-dominated galaxies and local density, mirrored in a positive correlation with disk + compact/classical bulge galaxies. \\

\noindent $\bullet$  At high stellar masses, we see significant increase in the elliptical fraction in high-density environments, which is mirrored in a decline in disk + diffuse/pseudo bulge and disk + compact/classical bulge. We propose that this suggests a second environmentally-driven process that transforms disk dominated systems into disk + compact/classical bulge galaxies, and both into ellipticals (major mergers and interactions?). This process leads to the observed morphology-density relation.               

\section*{Acknowledgements}

LJMD acknowledges support from the Australian Research Councils Future Fellowship and Discovery Project schemes (FT200100055 and DP250104611). ASGR, and SB acknowledge support from the Australian Research Council's Future Fellowship scheme (FT200100375). M.S. acknowledges support by the State Research Agency of the Spanish Ministry of Science and Innovation under the grants 'Galaxy Evolution with Artificial Intelligence' (PGC2018-100852-A-I00) and 'BASALT' (PID2021-126838NB-I00) and the Polish National Agency for Academic Exchange (Bekker grant BPN/BEK/2021/1/00298/DEC/1). This work was partially supported by the European Union's Horizon 2020 Research and Innovation program under the Maria Sklodowska-Curie grant agreement (No. 754510). Parts of this research were conducted by the Australian Research Council Centre of Excellence for All Sky Astrophysics in 3 Dimensions (ASTRO 3D), through project number CE170100013. DEVILS is an Australian project based around a spectroscopic campaign using the Anglo-Australian Telescope. DEVILS is part funded via Discovery Programs by the Australian Research Council and the participating institutions. The DEVILS website is \url{devils.research.org.au}. The DEVILS data are hosted and provided by AAO Data Central (\url{datacentral.org.au}).

\vspace{-5mm}

\section{Data Availability}

Data products used in this paper are taken from the internal DEVILS
team data release and presented in \cite{Davies21} and \cite{Thorne21}. These catalogues will be made public as part DEVILS
first data release described in Davies et al. (in preparation). GAMA data is taken from the GAMA DR4 public archive: \url{https://www.gama-survey.org/dr4/schema/}

\appendix

\section{Comparison between HST and JWST morphological classifications}
\label{app:JWST}

\begin{figure*}
\begin{center}
\includegraphics[scale=0.65]{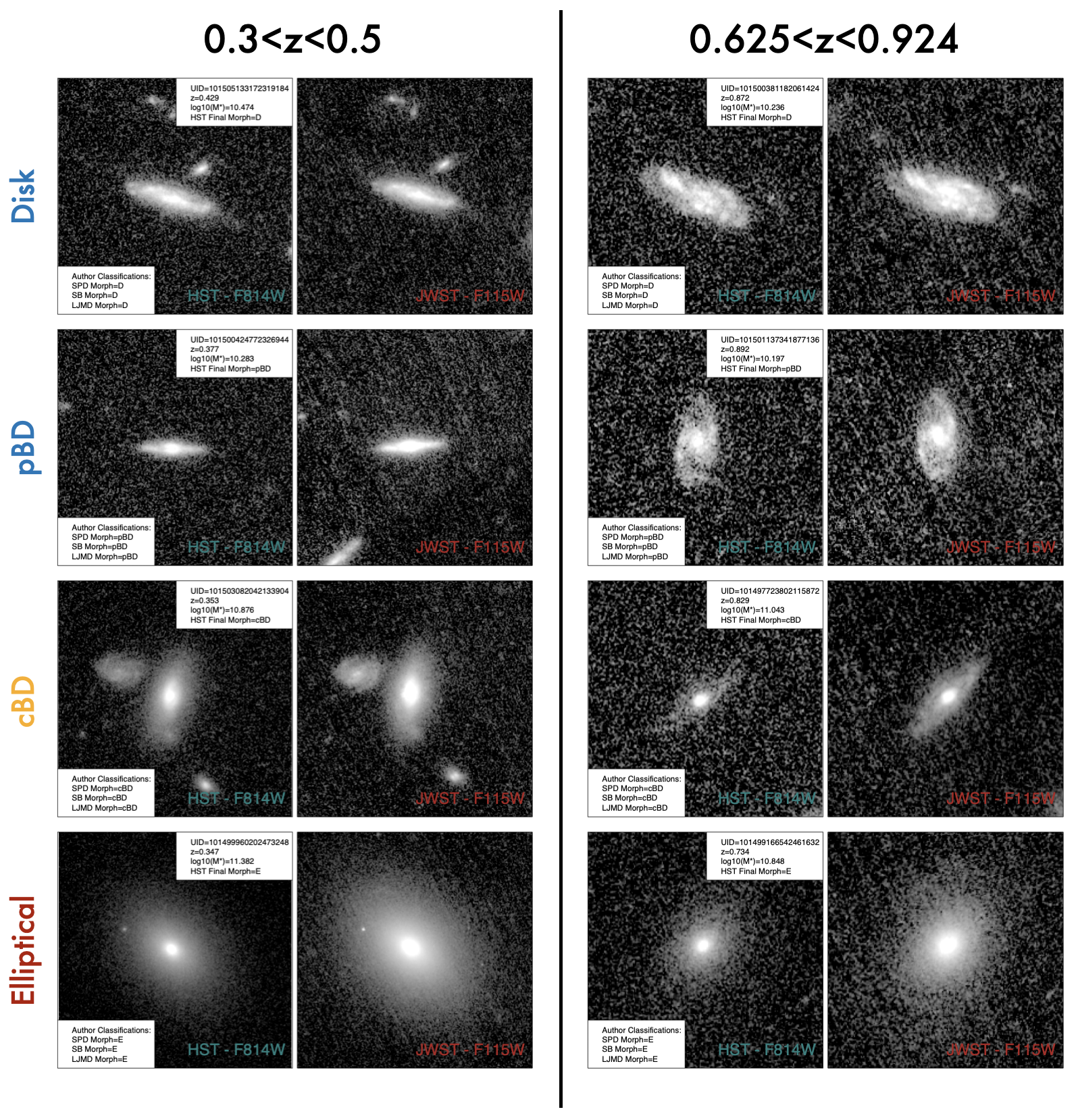}

\caption{Examples of JWST-F115W imaging for a random sample of galaxies in each morphological type. Left column shows galaxies in our low redshift spectroscopic only sample, while right column shows galaxies in our high redshift spectroscopic+photometric sample. In the top right legend in the left column we show DEVILS sample parameters for each source. In the bottom left legend we given individual classifier classifications from the \citet{Hashemizadeh21} work. While the JWST data reaches fainter surface brightness limits, this does not significantly impact morphological classifications at these stellar masses.    } 
\label{fig:JWSTlow}
\end{center}
\end{figure*}

\textcolor{black}{As discussed in Section \ref{sec:compVis}, a potential bias in the morphological classifications undertaken for DEVILS, in comparison to GAMA, is that the DEVILS-HST data is shallower in terms of surface brightness limits than GAMA-KiDs. We note in Section \ref{sec:compVis} that this is unlikely to cause issues for the sample used here, as our stellar mass limits impose a magnitude selection such that we only include galaxies with brightnesses far above the the data detection limits. However, as also noted in Section \ref{sec:compVis}, to confirm this, we compare our morphological classifications to deeper JWST COSMOS-Web imaging for a random subset of sources. Examples of comparisons between the HST imaging and JWST imaging for the low redshift spectroscopic-only sample (left column) and for the higher redshift sample (right column) are shown in Figure \ref{fig:JWSTlow}.  }

\section{Estimating errors on $\Sigma_{3}$ using the \textsc{Shark} semi-analytic model}
\label{app:densityErrors}

\textcolor{black}{Measurements of $\Sigma_{N}$ in observational samples can be biased by: i) chance projection effects when redshift precisions are poor, and ii) high peculiar velocities of systems leading to unknown true \textit{cosmological} redshifts, and therefore misidentification of the true closest $n^{th}$ neighbour. The former of these is likely not significant for our purely spectroscopic sample as redshifts have a high precision. For high-redshift sample, where photometric redshift are used, potential biases from projection effects are discussed in the following Appendix.} 

\textcolor{black}{However, in order to address the impact of peculiar velocities on our measurements of $\Sigma_{3}$, we explore the differences between true and observed $\Sigma_{3}$ values in the \textsc{Shark} semi analytic model \citep{Lagos24}. We use \textsc{Shark} simulated light cones that are produced for the WAVES survey, and have similar selection limits to DEVILS (and therefore also include GAMA-like galaxies). We then select samples from these light-cones matching the same stellar mass and redshift selections the GAMA and DEVILS spectroscopic samples used in this work. Within the \textsc{Shark} simulations they provide both \texttt{zobs} (the observed galaxy redshift including peculiar velocity) and \texttt{zcos} (the cosmological galaxy redshift not including peculiar velocity). First, using \texttt{zobs} we measure an observed $\Sigma_{3}$ for the \textsc{Shark} galaxies in an identical manner to our observations. Next, we measure the true physically separated $\Sigma_{3}$ using \texttt{zcos}. We then compare the standard deviation in log$_{10}$ offset between true and observed $\Sigma_{3}$ as a function of observed $\Sigma_{3}$. We find that this offset varies with $\Sigma_{3}$, but typically is of order 0.25-0.05\,dex. }

\begin{figure}
\begin{center}
\includegraphics[scale=0.55]{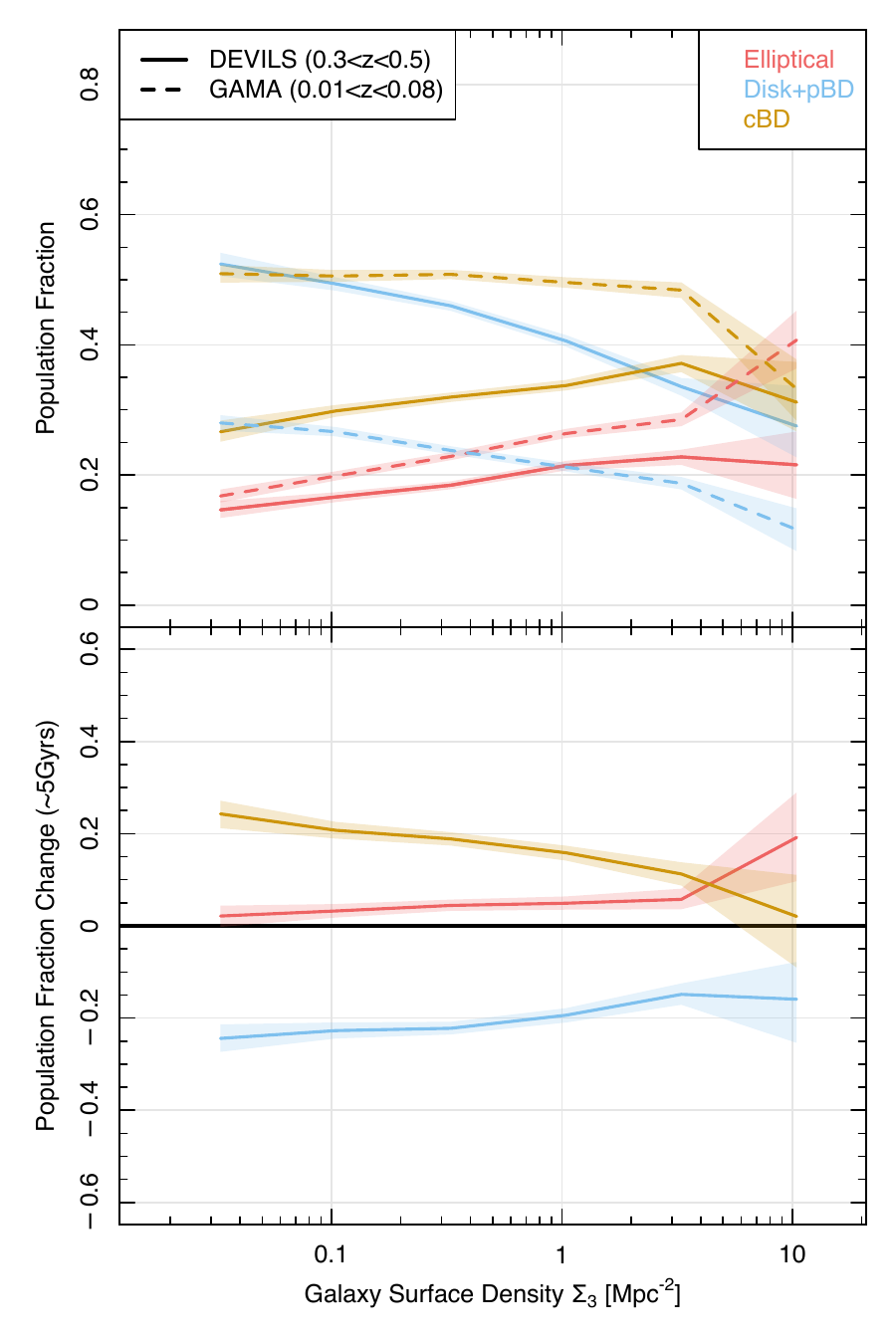}

\caption{The same as Figure \ref{fig:sigDense} but showing the impact of varying $\Sigma_{3}$, based of errors estimated using the \textsc{Shark} simulations (see text for details). Lines show the median of all monte-carlo realisations, while polygons show the interquartile range of the realisations. While some of the trends are reduced in significance, the overall trends remain.} 
\label{fig:sigErrors}
\end{center}
\end{figure}

\textcolor{black}{To determine how this impacts the results derived in this work. We produce 1000 monte-carlo realisations of the morphology density relation shown in Figure \ref{fig:sigDense}, but randomly scattering the observed $\Sigma_{3}$ values by the spread of possible offsets determined from \textsc{Shark} above ($i.e.$ at a given measured $\Sigma_{3}$ value we take the spread in true minus observed $\Sigma_{3}$ in \textsc{Shark} and randomly sample a new $\Sigma_{3}$ from that distribution). Figure \ref{fig:sigErrors} then shows the same as Figure \ref{fig:sigDense}, but now with lines showing the median morphology density relation derived from all monte-carlo realisations of the above process, and polygons showing the interquartile range of all realisations. While the significance of some of these trends are reduced, the overall relationships remain the same. This suggests that the results presented in this work are not significantly impacted by peculiar velocities leading to mis-calculations of $\Sigma_{3}$.}                                  

\section{Realisations Incorporating Photo-z Errors for DEVILS High-z sample}
\label{app:photoz}

\begin{figure*}
\begin{center}
\includegraphics[scale=0.44]{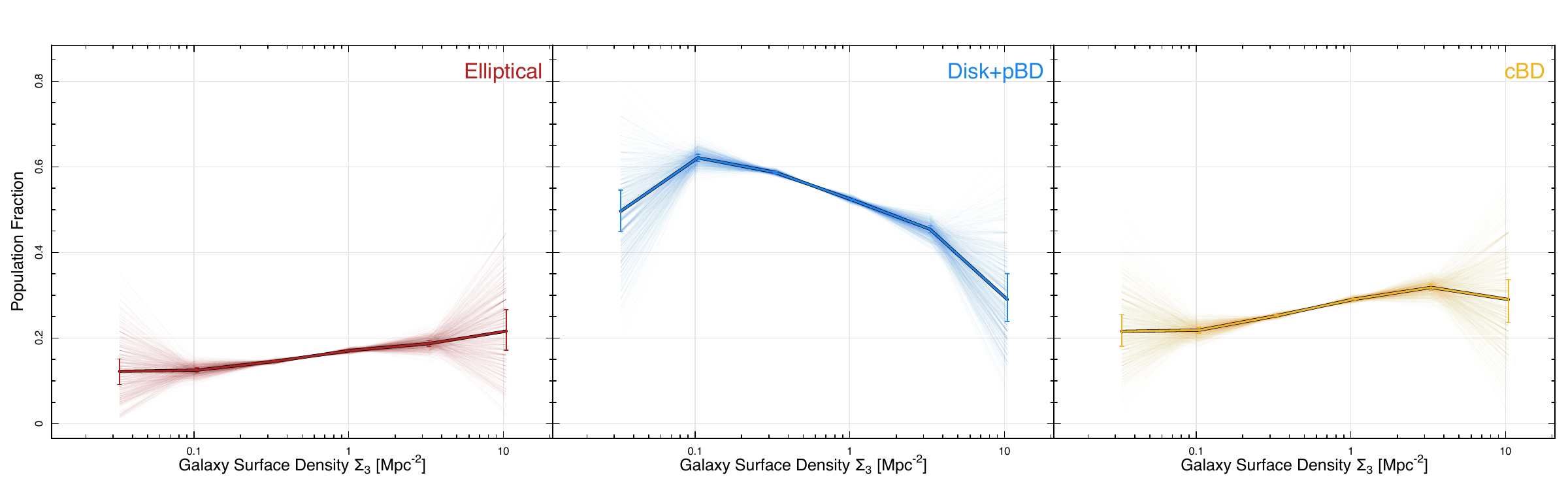}

\caption{The morphology-density relation for the DEVILS-highz sample for 1000 realisations of possible photometric redshifts. Each realisation is shown in a faint coloured line. The median relation is shown as the solid line. Errors show the $1\sigma$ spread of all realisations.  } 
\label{fig:shuffle}
\end{center}
\end{figure*}

As noted in Section \ref{sec:highz}, incorporating photometric redshifts into the sample can lead to biases in the measurement of $\Sigma_{3}$. Here we test the impact of imprecise photometric redshift measurements on our derived morphology-density relation and population fraction evolution. To do this we repeat the analysis in the paper using 1000 realisations of possible photometric redshifts for our DEVILS-highz sample. In each realisation, we identify all sources that have a photometric redshift. We then randomly sample a new photometric redshift from with a Gaussian distribution with $1\sigma$ width given by the photo-z $1\sigma$ errors (these come from the PAU and COSMOS2015 catalogues directly). Spectroscopic and grism redshifts remain unchanged. For these new photometric redshifts, we then re-calculate $\Sigma_{3}$, and then measure the DEVILS-highz morphological population fractions as a function of $\Sigma_{3}$. In Figure \ref{fig:shuffle} we then show the derived morphology-density relation for each realisation as faint coloured lines. The solid coloured line displays the median of all realisations and the error bars show the $1\sigma$ spread of all realisations. We see that at intermediate $\Sigma_{3}$ values the spread in realisations is very small, where the majority of the population lies. However, as we move to both low and high $\Sigma_{3}$ the spread increases for all samples - where single galaxies can have a stronger impact on the measurement of $\Sigma_{3}$. Firstly, this highlights the benefit of using purely spectroscopic samples to define local galaxy density (as we do in this work) as they are not encumbered by this potential source of error. We remind the reader that this sample still contains $\sim50\%$ spectroscopic redshifts, and even then the use of photometric redshifts can induce large biases in the measure of local galaxy density. This is why care must be taken when exploring the morphology-density relation using photometric redshifts alone. 

\begin{figure*}
\begin{center}
\includegraphics[scale=0.44]{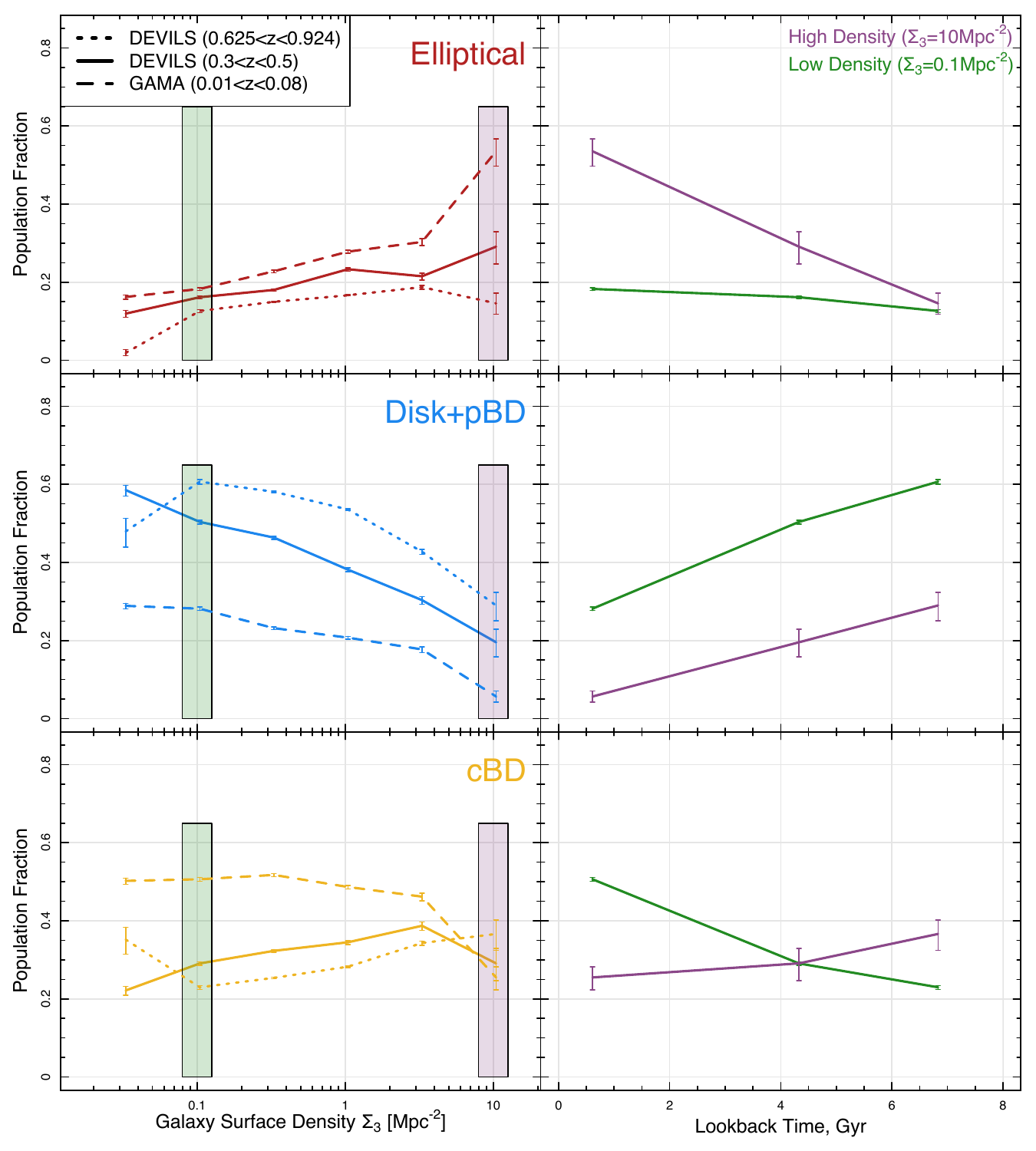}

\caption{The same as Figure \ref{fig:highz} but now using the median lines and errors from Figure \ref{fig:shuffle} for the DEVILS sample at $0.625<z<0.924$. While the population fractions marginally differer from those in Figure \ref{fig:highz}, the overall evolutionary trends remain the same.   } 
\label{fig:highz2}
\end{center}
\end{figure*}

Next, we reproduce Figure \ref{fig:highz} from the main body of the paper, but replace the DEVILS-highz data points with the median lines/errors from Figure \ref{fig:shuffle} - now shown in Figure \ref{fig:highz2}. We find that while incorporating photometric redshift errors into the analysis marginally shifts the population fractions at low and high $\Sigma_{3}$, the overall evolutionary trends seen in our work remain largely unchanged. As such, the photometric redshift uncertainties are not significantly impacting our results/conclusions.                   

\bsp	
\label{lastpage}
\end{document}